\documentclass[twocolumn]{aastex631} 

\usepackage{amssymb,amsmath,graphicx,natbib,hyperref}
\usepackage{multirow}
\usepackage{subfigure}
\usepackage{float}
\usepackage{footmisc}
\usepackage{color}
\usepackage[flushleft]{threeparttable}
\usepackage{enumitem}
\usepackage{natbib}
\usepackage{tabularx}
\usepackage{booktabs}

\newcommand{\asec}{$^{\prime\prime}$}
\newcommand{\kms}{\,km\,s$^{-1}$}

\newcommand{\um}{\,$\mu$m}

\newcommand{\vLSR}        {$v_{LSR}$}

\newcommand{\nuu}		  {$\nu$}

\newcommand{\s}		  {$\sim$}

\newcommand \percmsq {\,cm$^{-2}$}
\newcommand \percmcu {\,cm$^{-3}$}

\newcommand \co {$^{12}$CO}
\newcommand \thco {$^{13}$CO}

\newcommand{\cotwo}         {\mbox{CO$_2$}}

\newcommand{\hto}         {\mbox{H$_2$O}}
\newcommand{\chf}         {\mbox{CH$_4$}}
\newcommand{\oth}         {\mbox{O$_3$}}
\newcommand{\otwo}         {\mbox{O$_2$}}
\newcommand{\ntwoo}         {\mbox{N$_2$O}}
\newcommand{\notwo}         {\mbox{NO$_2$}}

\newcommand{\tex}		  {$T_\mathrm{ex}$}

\newcommand{\cc}		  {$^\circ$}
\newcommand{\wt}        {W3~IRS~5}

\shortauthors{Li et al.}

\shorttitle{MIR Water Absorption Lines in W3~IRS~5}

\begin{document}

\title{High-resolution SOFIA/EXES Spectroscopy of Water Absorption Lines in the Massive Young Binary \wt}

\correspondingauthor{Jialu Li}
\email{jialu@astro.umd.edu}

\author[0000-0003-0665-6505]{Jialu Li}
\affiliation{Department of Astronomy, University of Maryland, College Park, MD 20742, USA}

\author[0000-0001-9344-0096]{Adwin Boogert} 
\affiliation{Institute for Astronomy, University of Hawaii, 2680 Woodlawn Drive, Honolulu, HI 96822, USA}

\author[0000-0003-4909-2770]{Andrew G. Barr}
\affiliation{SRON Netherlands Institute for Space Research, Niels Bohrweg 4,
2333 CA Leiden, The Netherlands}

\author[0000-0002-6528-3836]{Curtis DeWitt}
\affiliation{SOFIA Science Center, USRA, NASA Ames Research Center, M.S. N232-12, Moffett Field, CA 94035, USA}

\author[0000-0002-3936-2469]{Maisie Rashman}
\affiliation{SOFIA Science Center, USRA, NASA Ames Research Center, M.S. N232-12, Moffett Field, CA 94035, USA}

\author[0000-0001-8341-1646]{David Neufeld}
\affiliation{Department of Physics and Astronomy, Johns Hopkins University, Baltimore, MD 20218, USA}

\author[0000-0001-8533-6440]{Nick Indriolo}
\affiliation{AURA for ESA, Space Telescope Science Institute, Baltimore, MD 21218, USA}

\author[0000-0001-8102-2903]{Yvonne Pendleton}
\affiliation{Department of Physics, University of Central Florida, Orlando, FL 32816, USA}

\author[0000-0003-2553-4474]{Edward Montiel}
\affiliation{USRA, SOFIA, NASA Ames Research Center, MS 232-11, Moffett Field, CA 94035, USA}

\author[0000-0002-8594-2122]{Matt Richter}
\affiliation{University of California Davis, Phys 539, Davis, CA 95616, USA}

\author[0000-0003-2029-1549]{J. E. Chiar}
\affiliation{Physical Science Department, Diablo Valley College, 321 Golf Club Road, Pleasant Hill, CA 94523, USA}

\author[0000-0003-0306-0028]{Alexander G. G. M. Tielens}
\affiliation{Department of Astronomy, University of Maryland, College Park, MD 20742, USA}
\affiliation{Leiden University, Niels Bohrweg 2, 2333 CA Leiden, The Netherlands}

\begin{abstract}
We present in this paper mid-infrared (5--8~\um) spectroscopy toward the massive young binary W3~IRS~5, using the EXES spectrometer in high-resolution mode ($R\sim$50,000) from the NASA Stratospheric Observatory for Infrared Astronomy (SOFIA). Many ($\sim180$) $\nu_2$=1--0 and ($\sim$90) $\nu_2$=2--1 absorption rovibrational transitions are identified. Two hot components over 500~K and one warm component of 190~K are identified through Gaussian fittings and rotation diagram analysis. Each component is linked to a CO component identified in the IRTF/iSHELL observations ($R=88,100$) through their kinematic and temperature characteristics. Revealed by the large scatter in the rotation diagram, opacity effects are important, and we adopt two curve-of-growth analyses, resulting in column densities of $\sim10^{19}$~\percmsq. In one analysis, the model assumes a foreground slab. The other assumes a circumstellar disk with an outward-decreasing temperature in the vertical direction. The disk model is favored because fewer geometry constraints are needed, although this model faces challenges as the internal heating source is unknown. We discuss the chemical abundances along the line of sight based on the CO-to-H$_2$O connection. In the hot gas, all oxygen not locked in CO resides in water. In the cold gas, we observe a substantial shortfall of oxygen and suggest that the potential carrier could be organics in solid ice.

\end{abstract}


\section{Introduction}
Massive stars reach the main sequence while deeply embedded and still accreting. While rich chemistry is driven as the central object warms and ionizes the environment, complicated physical activities such as accretion disks, outflows, shocks, and disk winds are involved \citep{beuther07, cesaroni07, zin07}. However,  the large distances to the observers, the high extinction at optical and near-infrared wavelengths, and the highly clustered environment impede a clear understanding of their formation and evolution processes.

High spectral resolution, pencil beam absorption line studies at mid-infrared (MIR) wavelengths provide a unique opportunity to probe the embedded phases in massive star formation \citep{lacy13}. First, the MIR continuum originates from the photosphere of the disk at a distance of tens to a few hundreds of au \citep{beltran16, frost21}. The effective pencil beam, therefore, avoids beam dilution issues that submillimeter observations are subject to. Second, molecules without dipole moments such as C$_2$H$_2$ and CH$_4$, which are among the most abundant carbon-bearing molecules, can only be observed through their rovibrational spectra in the infrared. Therefore, MIR spectroscopy at high resolution is a critical tracer of the physical conditions, the chemical inventory, and the kinematics of structures close to the massive protostars.

Among the rich chemical inventory in the regions associated with the protostars, water is of fundamental importance because it is one of the most abundant molecules in both the gas and ice phase. As its abundance varies largely between warm and cold gas \citep[e.g., ][]{draine83, kn96, bergin02}, water has a powerful diagnostic capability in probing physical conditions \citep{vds21}. However, due to its prevalence in the Earth's atmosphere, water is very difficult to observe from the ground. We present in this work the power of combining the Stratospheric Observatory for Infrared Astronomy \citep[SOFIA, ][]{young12} that flies observes between 39,000 and 45,000~feet and the Echelon Cross Echelle Spectrograph \citep[EXES, ][]{richter18} spectrometer  which resolves lines to several km/s to make the most of the diagnostic capability of water.

Previous studies of the MIR water absorption spectrum toward massive protostars~\citep{boonman03, boonman03a} have revealed a rich spectral content. High spectral resolution observations can provide much insight into the characteristics of these regions~{\citep[e.g.,][]{knez09, indriolo15, barr18, dungee18, indriolo18, rangwala18, goto19, barr20, indriolo20, nickerson21, barr22, barr22b, li22, nickerson23}}. Water has also been studied at high spectral resolution via pure rotational transitions, using the heterodyne instruments on board SWAS, Odin, and \textit{Herschel} Space Observatory~\citep[e.g.,][]{snell00, wilson03, char10, karska14}. Because of their limited spatial resolution, these observations mainly probed the large-scale environment of these sources.

We conducted high spectral resolution ($R$=50,000) spectroscopy from 5--8 $\mu$m with EXES on board SOFIA toward the hot core region close to the massive binary protostar \wt, which is is a luminous, massive protostellar binary in transition from the embedded to the exposed phase of star formation \citep{vdt00, vdt05}. {\wt\ is oriented along the northeast-to-southwest direction with a separation of 1.2\asec\ \citep[$\sim$2800~au at 2.3$^{+0.19}_{-0.16}$~kpc,][]{navarete19}. Following the nomenclature in \citet{vdt05}, we refer to the northeastern object of the binary as MIR1 and the southwestern one as MIR2. }

{Millimeter studies reveal complex structures surrounding the binary on a large scale (10$^3$--10$^5$~au), including a hot molecular core with outflows, jet lobes, shocks, and a circumbinary toroid~\citep{imai00, rodon08, wang12, wang13, purser21}. In the infrared $M$-band (4.7~$\mu$m), high spectral resolution ($R$=88,100) observations of rovibrational transitions of CO and its isotopologues~\citep{li22} show various absorbing structures along the pencil beam line of wight against the pencil beam MIR background. Physical conditions such as temperature and column density have been derived from these CO observations. While MIR1 and MIR2 are spatially resolved in \citet{li22}, the absorbing components are identified as a shared envelope ($\sim$50~K at $-38.5$~\kms), several foreground clumps produced by either J- or C-shocks (200--300~K from $-100$ to $-60$~\kms), and blobs that are likely associated with the circumstellar disks ($\sim$500~K from $-55$ to $-38.5$~\kms). The $M$-band study sets up the backdrop for understanding the origin of spatially unresolved water absorption lines in this paper, and we refer readers to Table 7 in \citet{li22} for a complete list of properties of each decomposed CO component.}

This paper presents the rich spectrum of rovibrational (the $\nu_2$ band) water lines observed in absorption toward \wt. The high spectral resolution spectroscopy allows us to separate and identify individual velocity components that are linked to different stars in the \wt\ binary and to derive the temperatures, the level-specific column densities as well as the total column densities (and/or the abundances). We describe our observations and data reduction in Section~\ref{sec:obs} and our analysis methods with both the rotation diagram and the curve-of-growth in Section~\ref{sec:da}. {We present in Section~\ref{sec:re} the properties of multiple dynamical components and set up the connection between these water components to CO components that are identified in \citet{li22}.} We discuss the chemical abundances along the line of sight based on the CO-to-H$_2$O connection in Section~\ref{sec:dis}.

\begin{deluxetable*}{ccc cccrrc ccr}[t]
\tablecolumns{12}
\tabletypesize{\scriptsize}
\tablecaption{Observational Parameters\label{table:obs}}
\tablehead{\colhead{Source} & \colhead{Date} & \colhead{Time} & \colhead{WN$_\textrm{cen}$} & \colhead{$\lambda$} & \colhead{$v_\textrm{geo}$} & \colhead{Long.} & \colhead{Lat.} & \colhead{Alt.}  & \colhead{ZA}& \colhead{Slit Height} & \colhead{PA} \\ &(UT) & (UT) & (cm$^{-1}$) & (\um)& (\kms) &(deg) &(deg)    &  (feet) & (deg)& (\asec) & \colhead{(\cc)} \\
 \colhead{(1)} & \colhead{(2)}  & \colhead{(3)} & \colhead{(4)} & \colhead{(5)} & \colhead{(6)} & \colhead{(7)} & \colhead{(8)}& \colhead{(9)} & \colhead{(10)} & \colhead{(11)}  & \colhead{(12)} }
\startdata
\wt&	2022-02-24	&	09:01:10 & 1841.2	&	5.36--5.51	&	-20.8	&	-117.11	&	49.11	&	41772	&	63.97	&	2.11	&	293.3--312.5	\\
&	2021-12-04	&	08:57:30	& 1794.3    &	5.48--5.67	&	-36.8	&	-124.41	&	29.93	&	43072	&	46.06	&	1.74	&	318.3--335.2	\\
&	2021-12-04	&	08:01:15	& 1740.7    &	5.65--5.84	&	-36.8	&	-132.92	&	26.80	&	43075	&	40.21	&	1.74	&	42.4--67.3	\\
&	2021-12-02	&	06:46:07	& 1688.1    &	5.83--6.02	&	-37.5	&	-91.80	&	34.73	&	41513	&	40.66	&	1.74	&	17.4--36.3	\\
&	2021-12-01	&	07:26:18	& 1637.4    &	6.01--6.20	&	-37.9	&	-104.60	&	38.89	&	40113	&	35.51	&	1.93	&	19.9--37.7	\\
&	2021-06-16	&	07:11:54	& 1630.8    &	6.01--6.20	&	-51.5	&	-98.82	&	45.37	&	42475	&	62.47	&	1.93	&	219.1--236.0	\\
&	2021-06-16	&	06:00:33	& 1592.7    &	6.18--6.37	&	-51.5	&	-113.11	&	48.40	&	41056	&	67.76	&	2.31	&	240.7--259.0	\\
&	2021-12-01	&	08:10:29	& 1586.1    &	6.19--6.37	&	-37.9	&	-98.06	&	41.39	&	40607	&	40.87	&	2.31	&	3.1--14.8	\\
&	2021-06-11	&	19:37:02	& 1544.2    &	6.35--6.61	&	-50.1	&	-111.17	&	48.82	&	41004	&	61.99	&	1.36	&	226.9--245.3	\\
&	2021-06-18	&	06:40:39	& 1543.9    &	6.35--6.61	&	-52.1	&	-100.56	&	45.11	&	41079	&	64.06	&	1.36	&	226.9--241.5	\\
&	2021-06-10	&	19:55:04	& 1488.6    &	6.59--6.85	&	-49.7	&	-98.07	&	45.88	&	42010	&	59.82	&	1.56	&	218.1--234.1	\\
&	2021-06-18	&	05:37:44	& 1488.6    &	6.59--6.85	&	-52.1	&	-114.09	&	47.81	&	41002	&	67.61	&	1.55	&	247.8--263.7	\\
&	2021-06-09	&	18:45:41	& 1444.5    &	6.79--7.06	&	-49.4	&	-123.32	&	51.26	&	41023	&	64.68	&	1.75	& 242.2--265.1	\\
& 2020-02-06\tablenotemark{a}   & 08:38:37  &1405.6   & 7.02--7.21 & -21.0 & -126.47 & 46.66 & 43007 & 55.00 & 3.06 & 318.7--331.3 \\ 
&	2022-02-24	&	07:53:55	& 1367.3    &	7.19--7.45	&	-20.8	&	-131.46	&	45.93	&	40589	&	57.87	&	2.33	&	318.3--335.2	\\
&2018-10-31\tablenotemark{b}	 & 03:26:20 & 1344.0   & 7.34--7.52 & -49.4 & -102.79 & 33.33 & 38020 & 39.28 & 3.25 & 152.7--173.0 \\ 
& 2020-02-07\tablenotemark{a} & 09:10:52 & 1318.8   & 7.49--7.68 & -21.0 & -152.16 & 49.01 & 41024 & 45.90 & 3.43 & 312.5--341.6 \\ 
&	2021-06-09	&	20:06:39	& 1283.3    &	7.67--7.92	&	-49.4	&	-107.06	&	48.97	&	43011	&	59.73	&	2.51	&	226.3--238.9	\\
Sirius &	2022-02-24	&	06:45:01	& 1367.3 &	7.18--7.46	&	--	&	-135.36	&	41.32	&	39098	&	62.13	&	--	&	--\\
\enddata
\tablenotetext{a}{{Archival data} (AOR\_ID: 07\_0063).}
\tablenotetext{b}{Archival data (AOR\_ID: 76\_0004).}
 \tablecomments{Column (1): Sirius is the standard star for the observation session on 2022-02-24 at 7.19--7.45~\um. Column (4): the central wavenumber of the setting. Column (6): the Earth's Doppler velocity with respect to \wt\ at the time of the observation. Column (7)--(10): the longitude, latitude, altitude, and zenith angle of the telescope of the observation session. Column (11)--(12): the height and the position angle (from north to east) of the slit. Note that the shortest slit height (1.36\asec) is longer than the distance between the binary of \wt~(1.2\asec). For all settings, the slit width is fixed to 3.2\asec.}
\end{deluxetable*} 

\begin{figure*}
    \centering
    \includegraphics[width = \linewidth]{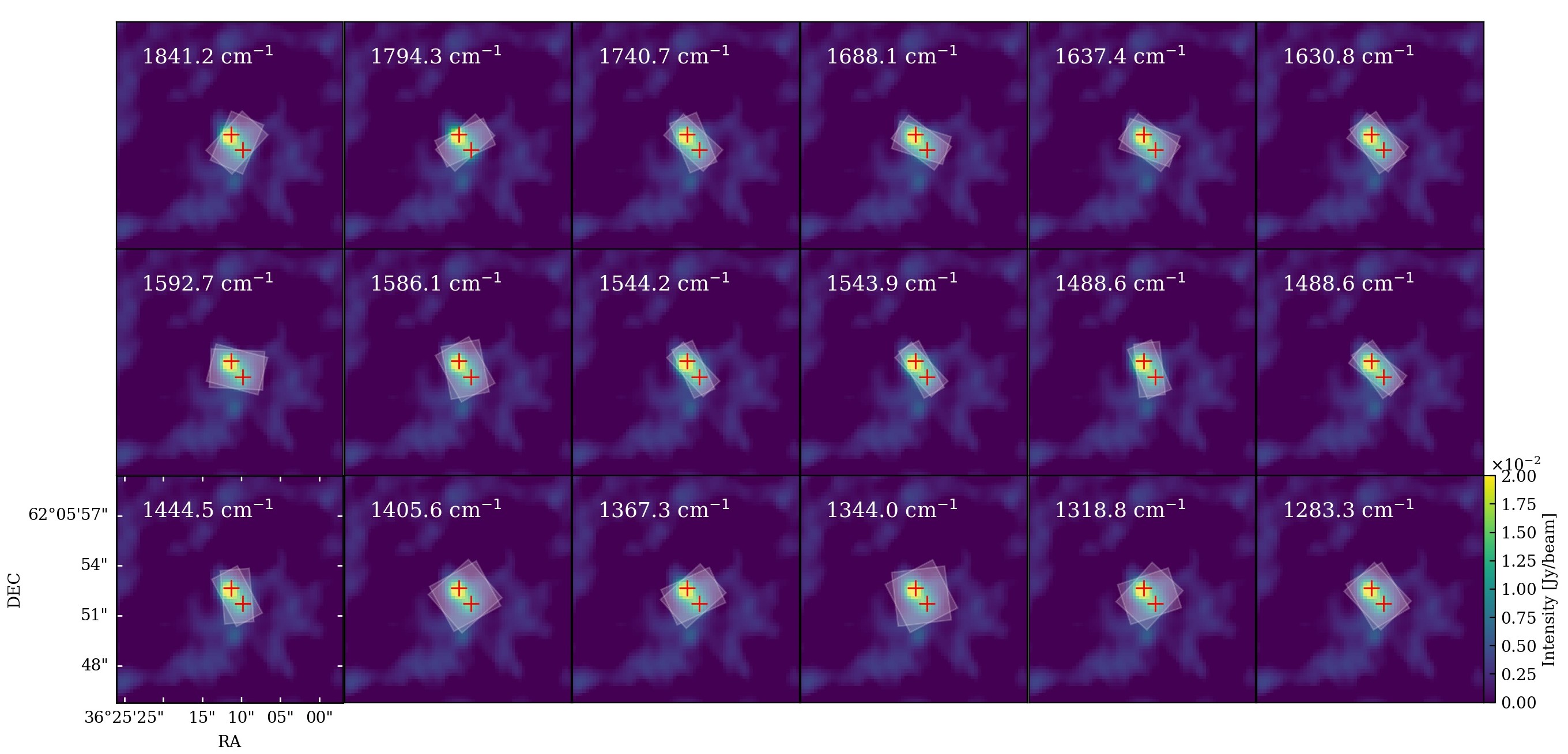}
    \caption{Slit coverage (white rectangles) on top of \wt\ at different observational settings. Only the initial and final positions are plotted in the figure. The range of the position angles is listed in Table~\ref{table:obs}. The three settings centered at 1405.6, 1344.0, and 1318.8~cm$^{-1}$ are the archival data. The background is the PdBI 3.4~mm image of \wt\ adopted from \citet{rodon08}, and the two red crosses \citep[RA: 02 25 40.68, DEC: 62 05 51.53 and RA: 02 25 40.78, DEC: 62 05 52.47,][]{vdt05} mark the positions of the proto-binary stars.  }
    \label{fig:slit-pos}
\end{figure*}

\section{Observations and Data Reduction}\label{sec:obs}

Object \wt\ was observed with the EXES spectrometer
\citep{richter18} on board SOFIA observatory from 2021 June to 2022 February as part of programs 08\_0136, 09\_0072. Archival data observed in programs 07\_0063 and 76\_0004 were also included in this study. The full spectral survey covers a wavelength range of 5.3--7.9~\um\ in 18 observational settings and was observed under the HIGH-LOW cross-dispersed mode. The observational parameters of all 18 settings are listed in Table~\ref{table:obs}. For all settings, the slit width is fixed to 3.2\asec\  to limit slit losses perpendicular to the slit at a SOFIA {PSF (point spread function) FWHM (full width at half maximum)} of $\sim$3.0--3.5\asec, which provides a spectral resolution of $R\sim50,000$, or equivalently, a velocity resolution of 6~\kms. The slit length is dependent on the wavelength and the angle of the echelle grating and is in the range of 1.36--{3.43}\asec\ after accounting for the anamorphic magnification (Figure~\ref{fig:slit-pos}). Off-slit nodding was applied to remove the background night sky emission and the telescope thermal emission.  

The EXES data were reduced with the SOFIA Redux pipeline \citep{clarke15}, which has incorporated routines originally developed for the Texas Echelon Cross Echelle Spectrograph \citep{lacy03}. The science frames were despiked and sequential nod positions subtracted, to remove telluric emission lines and telescope/system thermal emission. An internal blackbody source was observed for flat fielding and flux calibration and then the data were rectified, aligning the spatial and spectral dimensions. The wavenumber solution was calibrated using sky emission spectra produced for each setting by omitting the nod-subtraction step. We used wavenumber values from HITRAN~\citep{rothman13} to set the wavelength scale. The resulting wavelength solutions have 1$\sigma$ errors of 0.5~\kms.

Before measuring intrinsic lines in the source, the spectra had to be corrected for telluric absorption. {The telluric absorption lines can be separated from lines that are from W3~IRS~5 thanks to the Doppler shifts (Table~~\ref{table:obs}).} Ideally, {removing telluric absorption} is done by taking a spectrum of a featureless hot star immediately before or after the target observation. This method also has the advantage of removing instrumental baseline effects, such as fringing, present in present in both spectra. Unfortunately, it was impractical to schedule calibrators for every part of the W3~IRS~5 survey due to the limitations of airborne observation scheduling and the scarcity of bright standard stars. Only one survey setting was observed with the adjacent calibrator star Sirius (7.28--7.46~\um, Table~\ref{table:obs}). {Thus we relied on atmospheric transmittance models created with PSG~\citep[Planetary Spectrum Generator, ][]{villa18} and tuned the {H$_2$O} column and pressure for the longitude, latitude, and altitude of the observations (Appendix~\ref{app:tl}) to remove telluric features, including minor features that may overlap with water lines from \wt. The quality of these corrections was verified by telluric lines not overlapping with W3IRS5 lines. To establish the reliability of the construction procedure of the PSG models, we compared the Sirius spectrum with the PSG model built from 7.28--7.46~\um~(see Appendix~\ref{app:sirius}). We conclude that the telluric correction yields uncertainties on the equivalent widths at
the level of 10\%. After the PSG models are divided, further data reduction such as removing the local baseline is still conducted.}

Since the distance between the binary (1.2\asec) is smaller than the SOFIA PSF ($\sim$3--3.5\asec), \wt\ is not spatially resolved in the observed spectra. As both sources contribute to the observed flux in our slit, the derived equivalent width of an absorption associated with one IR source will be diminished by the continuum emission from the other source. The two sources have very comparable MIR brightness \citep[see the SpeX observations in][]{li22} and we include a factor {of} 2 in our analysis to account for this effect throughout this study unless otherwise stated. {In other words, we think the relative absorption depth should be two times deeper because the continuum level should be two times weaker.} It should be mentioned that the actual contribution of the non-absorbing source will depend on the slit orientation over the source and this changes slightly between the different grating settings (see Figure~\ref{fig:slit-pos}). We have decided to accept this additional source of uncertainty to the results in view of the uncertain brightness distribution of the two sources on the sky.

\section{Data Analysis}\label{sec:da}

\begin{figure*}[!t]
    \centering
    \includegraphics[width = 0.97\linewidth]{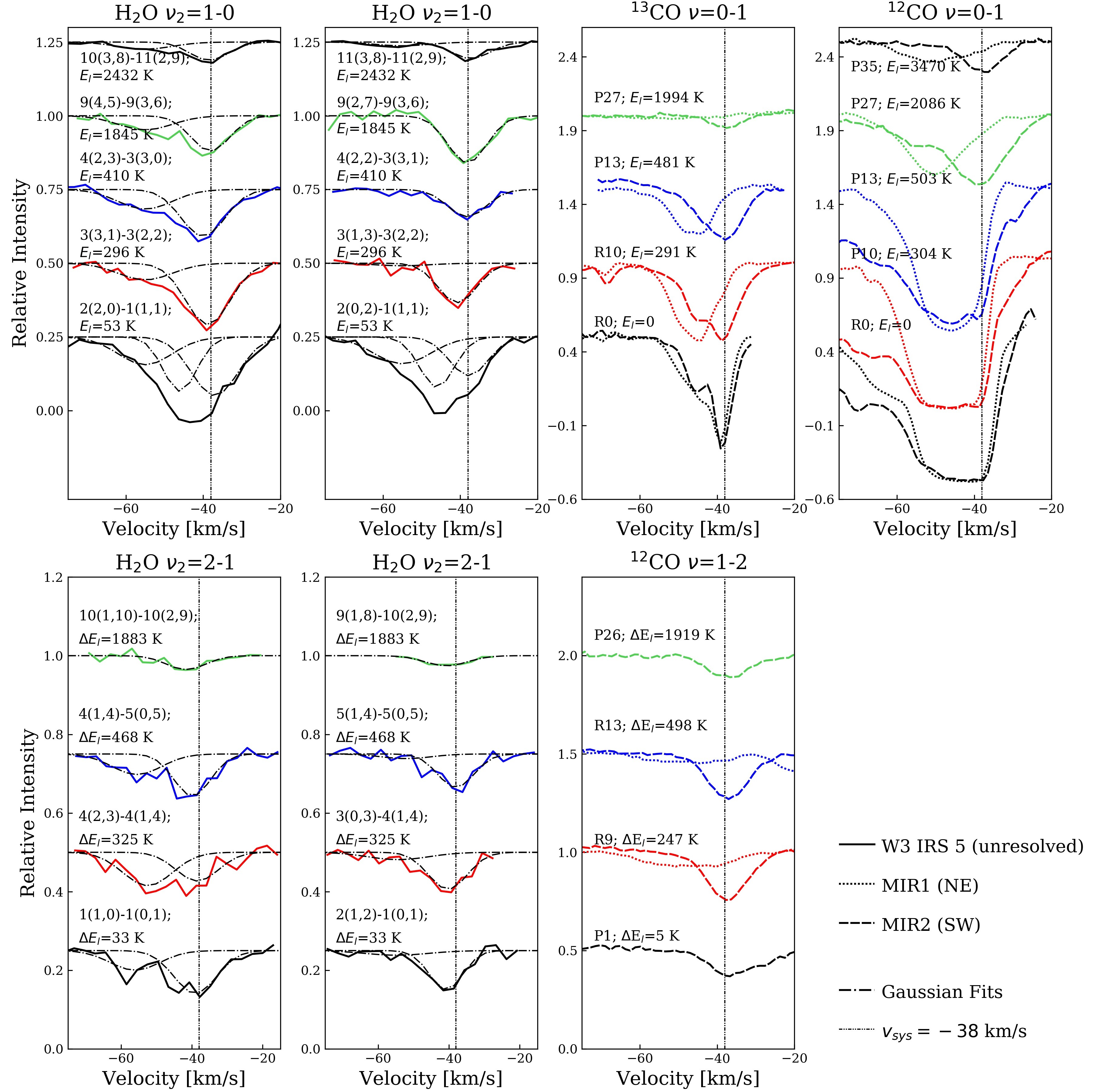}
    \caption{Selected H$_2$O $\nu_2$=1--0, H$_2$O $\nu_2$=2--1 (from this study), \thco~\nuu=0--1, \co~\nuu=0--1, and \co~\nuu=1--2 \citep[adopted from ][]{li22} absorption lines observed toward \wt. This binary, consisting of sources MIR1 and MIR2, is spatially resolved in ground-based IRTF studies of CO and their observed  line profiles are indicated by dotted and dashed lines. MIR1 and MIR2 are not spatially resolved in the {H$_2$O} observations on SOFIA. However, the high spectral resolution of these observations assists in the disentanglement of MIR1 and MIR2. The dashed vertical lines at $-38$~\kms\ are the systematic velocity. Across the panels, transitions with similar energy levels are represented by the same color. For vibrationally excited states, $\Delta E_l$ is the energy difference relative to the ground state energy of $\nu_2=1$ for water or $\nu$=1 for CO. Distinct kinematic components in H$_2$O transitions are present under different excitation conditions. Gaussian fitting profiles on top of H$_2$O lines are centered at $-54.5$ (`H2'), $-45$ (`W'), and $-39.5$ (`H1')~\kms~(see \S~\ref{sec:da} for the nomenclature). We note that H$_2$O line profiles can originate from states with comparable energies and yet may differ significantly in their line profiles due to the opacity effect (see \S~\ref{subsec:bd}).}
    \label{fig:compare}
\end{figure*}

{By comparing with the spectra constructed via the existing laboratory line information~\citep[HITRAN,][]{rothman13} and LTE models}, we identified over 180 H$_2$O $\nu_2$=1--0 and about 90 H$_2$O $\nu_2$=2--1 absorption lines in this survey. As shown in Figure~\ref{fig:compare}, the velocity ranges of the absorbing components are very similar to those present in \thco\ or in high-J \co\ lines~\citep{li22}. All components are blueshifted compared to or are located at the cloud velocity \vLSR$=-38$~\kms~\citep{vdt00}. Each kinematic component is characterized by different excitation conditions.

We fit the $\nu_2$=1--0 absorption line profiles with a sum of multiple Gaussians indicated as the $i$th component in the form of 
\begin{equation}
    \frac{I_\textrm{obs}(v)}{I_c} = 1 - \sum_i \frac{W_i}{\sigma_{v, i}\sqrt{2\pi}}\textrm{exp}\left(-\frac{1}{2}\frac{(v - v_{{LSR}, i})^2}{\sigma_{v, i}^2} \right) \label{equ:gaussian}
\end{equation}
using the \texttt{curve\_fit} function in \texttt{scipy}. {In Equation~\ref{equ:gaussian}, $I_\textrm{obs}(v)$ is the observed line intensity profile in the velocity space, $I_c$ is the intensity of the continuum, $W_i$ is the area or the equivalent depth of the Gaussian component, $\sigma_{v, i}$ is the standard deviation, and $v_{\textrm{cen}, i}$ is the center velocity of the component.} The fitting results are determined by restricting the range of the free parameters. Specifically, two components centered at $\sim-39.5$~\kms\ and $-54.5$~\kms\ are present across all energy levels, while the component centered at $-45$~\kms\ only appears in transitions originating from relatively low energy states ($<$800~K).  {As we will explain in Section~\ref{subsec:bd}, we name the three components at $\sim-39.5$~\kms, $-45$~\kms, and $-54.5$~\kms\  as `H1', `W', and `H2'.} For very low energy states ($<$200~K), an extra component at $-39.5$~\kms\ is possibly needed for a better fitting result (see Appendix~\ref{section:hidden}), and we discuss its existence and the implications in \S~\ref{subsec:hot}. 

We therefore fit two and three Gaussians above and below 800~K, separately. Above 800~K, we constrain the velocity center to $-43$ to $-36$~\kms\ for the `H1' component, and the `H2' component with a relative velocity of $-17$ to $-14$~\kms\ guided by an initial guess. The line widths, $\sigma_v$, were constrained to $4$--$6$ and $7$--$9$~\kms. We apply the same constraints to the $\nu_2$=2--1 absorption line profiles. 
{On the other hand, we think the profiles of the low energy transitions ($<$800~K) are a blend of more than two components, including the rather weak {`W' and `H2'} components. After a careful assessment, we decided to limit the number of free parameters in the fit to these transitions as, from a physical perspective, we expect that transitions from a given temperature component will occur at consistent central velocities.  We keep using a range for the velocity dispersion in the fitting procedure as we did for the high energy transitions. More specifically, for the low energy transitions, we fix the {rather weak} `W' and `H2' with the central velocity, {\vLSR}, of $-45$ and $-54.5$~\kms, and the velocity dispersion parameter, $\sigma_v$ from $4$--$5$~\kms.} {We note that the H2 ($-54.5$~\kms) component disappears in some of the low-energy transitions (see the second panel in Figure~\ref{fig:compare}) as a result of the opacity effect (see \S~\ref{subsec:bd}).} In this case, we only fit the line profile with one Gaussian for the `H1' component. 
We list in Appendix~\ref{app:table} in detail the parameters of each individual component such as the central velocity, velocity width, and equivalent widths, and {report `H1' as $-39.1\pm2.4$~\kms, and `H2' ($>$800~K) as $-54.1\pm3.0$~\kms, respectively.}

\subsection{Rotation Diagram Analysis and the Opacity Problem}\label{subsec:bd}

After the distinct velocity components are determined, we use a rotation diagram to provide a first view of their characteristics. If the absorption lines are optically thin, we can get the column density $N_l$ in the lower state of a transition directly from the integrated line profile by
\begin{equation}
N_l = 8\pi/(A_{ul} \lambda^3) g_l/g_u \int \tau(v)dv, \label{eq:1}
 \end{equation}
in which {$\lambda$ is the wavelength}, $A_{ul}$ is the Einstein A coefficient, $g_l$ and $g_u$ are the statistical weight of the lower and upper levels, {$v$ is the velocity}, and
\begin{equation}
\tau(v) = -\textrm{ln}(I_v/I_c),\label{eq:2}
\end{equation}
where $I_v$ and $I_c$ are {the intensity of the absorption line and the continuum, and $I_v$ of each identified velocity component in a given transition is derived from the Gaussian fitting parameters~(equation~\ref{equ:gaussian})}. All spectral line parameters used in this study are adopted from HITRAN \citep{rothman13}.

If the foreground absorbing gas is in LTE, the population of one rotational level can be described with the Boltzmann equation~{\citep{gl99}},
\begin{equation}
    \frac{N_l}{g_l} = \frac{N_\textrm{tot}}{Q(T_\textrm{ex})} \textrm{exp}\left(-\frac{\Delta E_l}{k_B T_\textrm{ex}}\right), \label{eq:bd}
\end{equation}
in which \tex\ is the excitation temperature, $N_\textrm{tot}$ is the total column density, {$\Delta E_l$ is the relative energy 
between the excitation state and the ground state energy of the vibration state, and equals to $E_l$ for $\nu_2$=1--0 transitions\footnote{For $\nu_2$=2--1 transition, $\Delta E_l$ is relative to the ground state energy of $\nu_2=1$ ($J_{K_a, K_c}=0_{0,0}$, 2294.7~K).}.} {$Q(T_\textrm{ex})$ is the partition function, and $k_B$ is the Boltzmann constant}.  {Specifically, ln($N_l/g_l$) and $E_l/k_B$ of all absorption lines constructs the so-called rotation diagram. For a uniform excitation temperature, ln($N_l/g_l$) and $E_l/k_B$ fall on a straight line.} The inverse of the slope represents the temperature and the intercept represents the total column density over the partition function.

\begin{figure*}
    \centering
    \includegraphics[width=\linewidth]{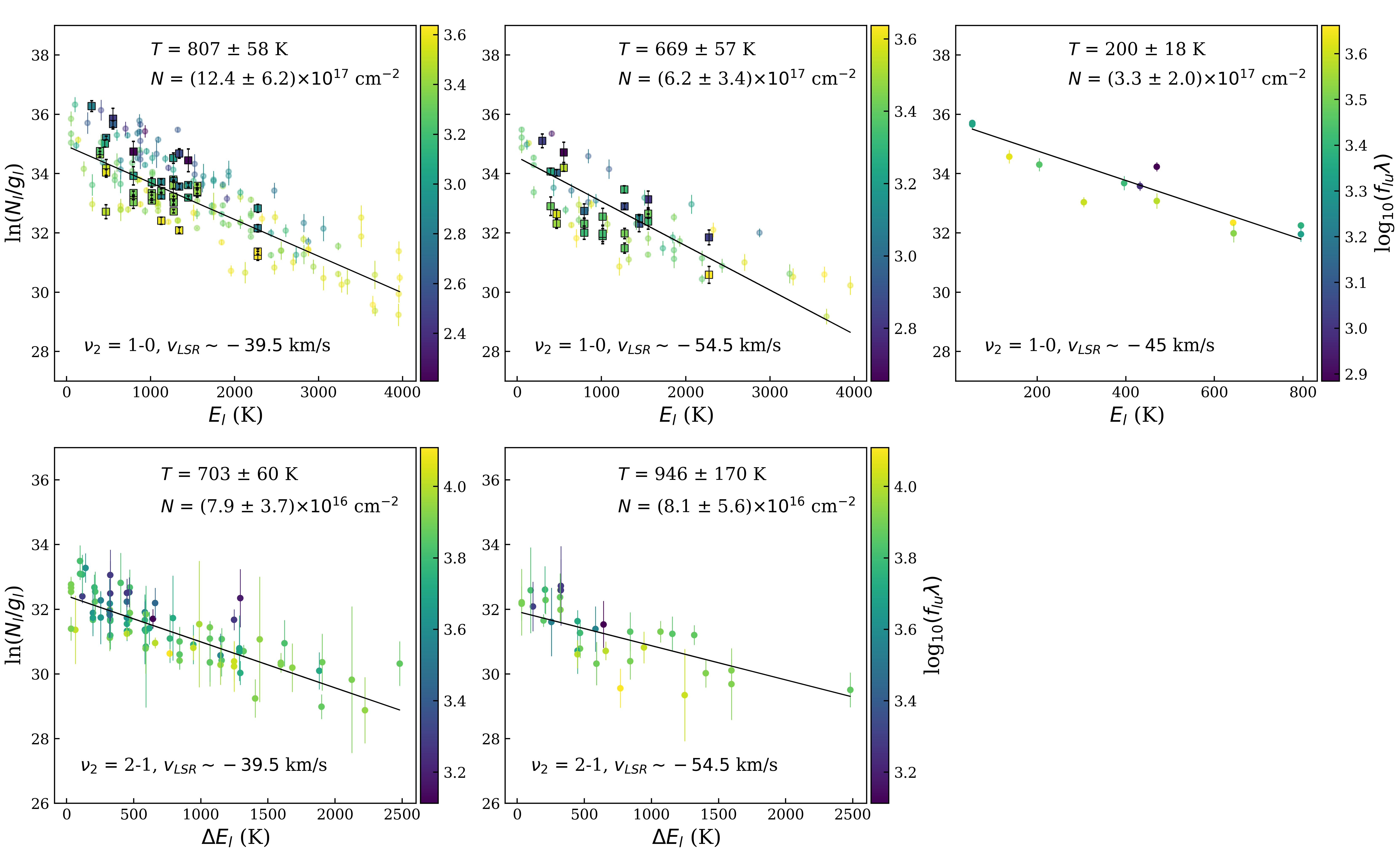}
    \caption{Rotation diagrams of H$_2$O $\nu_2$=1--0 (\textit{upper panels}) and $\nu_2$=2--1 (\textit{lower panels}) transitions from the decomposed velocity components: $-39.5$ (`H1'), $-45$ (`W'), and $-54.5$ (`H2')~\kms. For $\nu_2$=2--1 transitions, $E_l$ in the $x$-axis are $\Delta E_l$ relative to the ground state energy of $\nu_2=1$ ($J_{K_a, K_c}=0_{0,0}$, 2294.7~K). Square data points represent a collection of lines that share the same lower level as shown in Figure~\ref{fig:nl}. The $N_l$ are derived from equation~\ref{eq:1} with Gaussian fitting on the line profiles assuming optically thin absorption. The color code is log$_{10}(f_{lu}\lambda)$, which is representative of the intrinsic strength (see \S~\ref{subsec:cog}). The parameter $f_{lu}$ is the oscillator strength and $\lambda$ is the wavelength.}
    \label{fig:bd}
\end{figure*}

We present the rotation diagrams of the rovibrational \hto\ lines from \wt\ in Figure~\ref{fig:bd}. For the three velocity components centered at $-39.5, -45$, and $-54.5$~\kms\ identified in $\nu_2$=1--0 transitions, the derived temperatures are {{807$\pm$58, 200$\pm$18, and 669$\pm$57~K}} (Table~\ref{table:sum-prop}). {The three components are hence named `H1', `W', and `H2', in which `W' and `H' stands for ``warm" and ``hot" following the nomenclature in \citet{li22}.} {For $\nu_2$=2--1 transitions, only the component `H1' and `H2' at $-39.5$ and $-54.5$~\kms\ are identified, and the derived temperatures are 703$\pm$60 and 946$\pm$170~K (Table~\ref{table:sum-prop})}\footnote{The derived temperature of `W' of $\sim$200~K from the $\nu_2$=1--0 transitions explains the non-detection of `W' in the $\nu_2$=2--1 transition. According to equation~\ref{eq:tvib}, the column density of `W' is expected to be three orders of magnitude weaker than that of `H1' and `H2' in $\sim$600~K.}.

The derived total column densities of `H1', `W', and `H2' in the $\nu_2$=0 state are (1.2$\pm$0.6)$\times10^{18}$, (3.3$\pm$2.0)$\times10^{17}$, and (6.2$\pm$3.4)$\times10^{17}$~\percmsq, respectively (Table~\ref{table:sum-prop}). However, it is noticeable that the large scatter in the rotation diagram exceeds what the error bars can account for in components H1 and H2. 
Specifically, as shown in Figure~\ref{fig:bd}, for transitions that share the same lower state (and the same $g_l$), the difference in the derived $N_l$ is up to an order of magnitude. 

\begin{figure*}
    \centering
    \includegraphics[width=\linewidth]{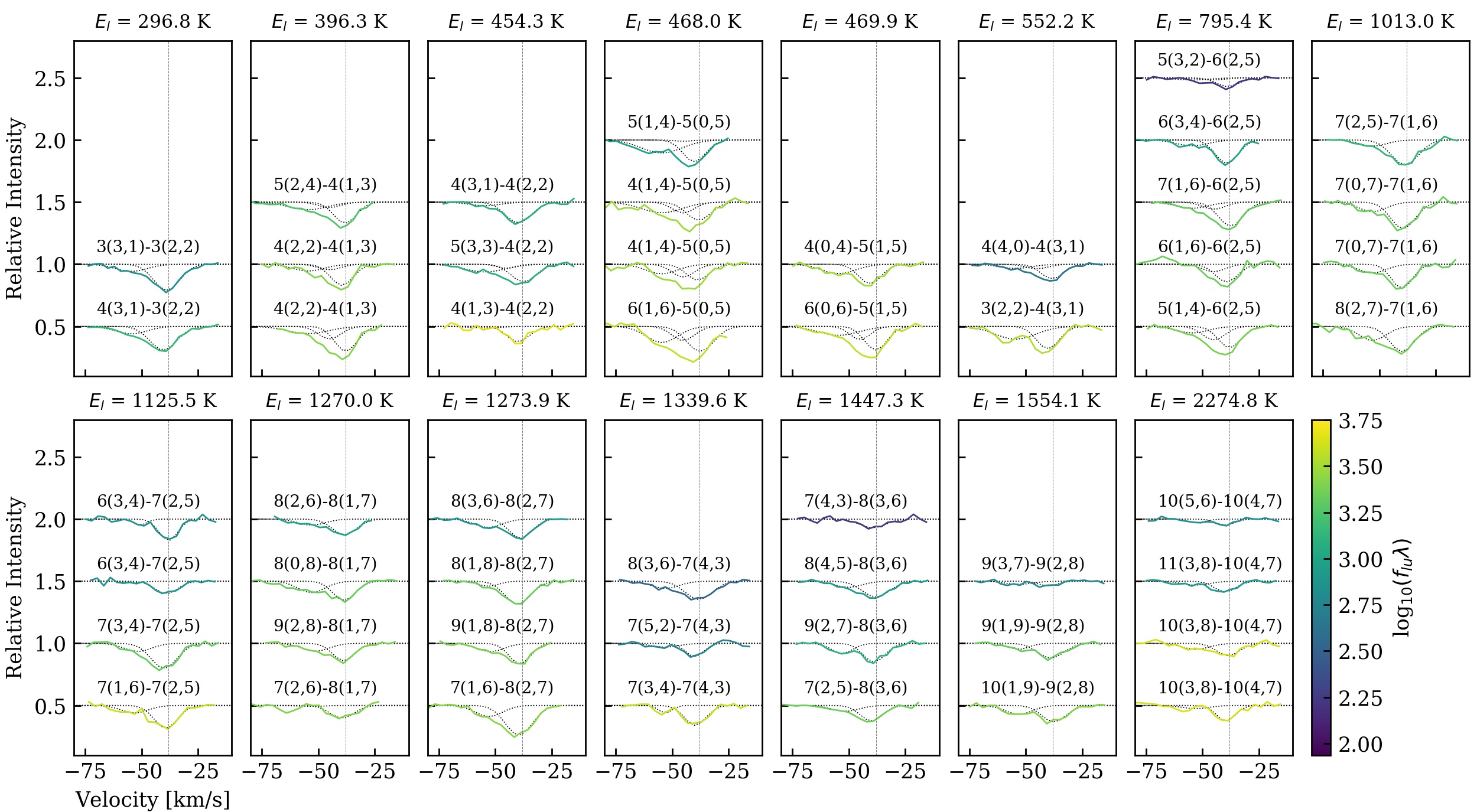}
    \caption{Selected \hto\ $\nu_2$=1--0 absorption lines observed toward \wt. In each panel, all rovibrational lines share the same lower energy state, and thus column density. But, while the intrinsic strength, log($f_{lu}\lambda$), increases from top to bottom, the equivalent width often does not, indicating that the lines are optically thick. Some transitions appear in more than one setting and result in duplicated entries in the figure. The line profiles are color-coded following the color bar in Figure~\ref{fig:bd} (see Section~\ref{subsec:cog}), and are listed in each panel from top to bottom with increasing log$_{10}(f_{lu}\lambda)$, which is representative of the opacity. Gaussian fitting {in dotted lines} for `H1' and `H2' components at $E_l>800$~K (for `W', `H1', and `H2' at $E_l<$800~K) are presented together with each absorption line. The dashed vertical lines at $-38$~\kms\ indicate the systematic velocity of \wt~\citep{vdt00}.}
    \label{fig:nl}
\end{figure*}

We can further illustrate this problem by directly comparing the line profiles of the transitions that have a common lower energy level, {and present such a comparison in Figures~\ref{fig:compare} and \ref{fig:nl}. As shown in Figures~\ref{fig:compare} and \ref{fig:nl}, lines that share the same lower level but different oscillator strength ($f_{ul}$) or Einstein A do not necessarily have the same absorption intensity and/or the equivalent width, } which is defined as
\begin{equation}
    W_\nu = \int (1 - I_\nu/I_c) d\nu \label{eq:ew1}
\end{equation}in the frequency space. \textcolor{black}{This behavior where transitions with high Einstein A's fall systematically below the relation provided by transitions with low Einstein A's in the rotation diagram (Figure~\ref{fig:bd}) is characteristic for opacity effects. Specifically, when transitions are optically thick, an increase in absorption strength (due to an increase in Einstein A) results in only a small increase in line width (not in depth) and hence marginally increase the equivalent width. This effect was noted earlier in \citet{indriolo20} and \citet{barr22} in studies of MIR \hto\ rovibrational lines toward massive protostars AFGL~2136 and AFGL~2591.} 

\subsection{Curve-of-growth Analysis}\label{subsec:cog}

Considering that $\tau(v)$ in equation~\ref{eq:1} does not represent an optically thin Gaussian core, the definition of the equivalent width in equation~\ref{eq:ew1} can be written as
\begin{equation}
  \begin{aligned}
   W_\nu &= \int (1 - I_\nu/I_c) d\nu \\ &= \int (1 - \textrm{e}^{-\tau(\nu)}) d\nu \\ &= \int  (1 - \textrm{e}^{-\tau_p H(a, v)}) d\nu 
 \end{aligned}
\end{equation}
{in which the integration is over the frequency $\nu$. The Voigt profile $H(a, v)$ is defined as \citep[equation 9-34 in][]{mihalas78}:  
\begin{equation}
    H(a, v) = \frac{a}{\pi}\int_{-\infty}^{+\infty} \frac{e^{-y^2}dy}{(v-y)^2 + a^2}. \label{eqn:voigt}
\end{equation}
The parameter, $v$, is defined as
\begin{equation}
    v = \frac{\nu - \nu_0}{\Delta \nu_D}
\end{equation}
and represents the shift from the line center in Doppler units. $\Delta \nu_D$ is the Doppler width in frequency space.} The parameter $a$ is the damping factor. The peak optical depth $\tau_p$ is

\begin{equation}
    \tau_p = \frac{\sqrt{\pi}e^2}{m_e bc}N_{l}f_{lu}\lambda. \label{eq:drain-tau}
\end{equation}
In equation~\ref{eq:drain-tau}, $e$ is the electron charge, $m_e$ is the electron mass, {$c$ is the speed of light,} and $f_{lu}$ is the oscillator strength. The Doppler parameter in velocity space, $b$, is related to the full width at half maximum of an optically thin line by $\Delta v_{\textrm{FWHM}} = 2\sqrt{\textrm{ln}2}b$.

Equation~\ref{eq:drain-tau} clearly shows that for lines that share the same lower level and have the same $N_l$, a difference in $f_{lu}\lambda$ will lead to different $\tau_p$. Defining log$_{10}(f_{lu}\lambda)$ as the representative for the line strength, lines with a larger line strength have larger equivalent widths (as shown in Figure~\ref{fig:nl}), and as a result the $N_l$ derived via equation~\ref{eq:1} will be underestimated (as shown in Figure~\ref{fig:bd}).

\subsubsection{Slab Model of a Foreground Cloud}\label{subsec:sm}

A curve-of-growth analysis \citep{rodgers74} is required to reconcile the opacity problem and to correctly derive $N_l$ and $N_\textrm{tot}$~{\citep[equation 9.27 in][]{draine11}}:

\begin{multline}
    \frac{W_\lambda}{\lambda}  \approx \\
        \begin{cases}
      \frac{\sqrt{\pi}b}{c}\frac{\tau_p}{1+\tau_p/(2\sqrt{2})}&  \text{for} ~\tau < 1.254\\
      \frac{2b}{c}\sqrt{\textrm{ln}(\tau_p/\textrm{ln}2) + \frac{\gamma \lambda}{4b\sqrt{\pi}}(\tau_p - 1.254)} &  \text{for} ~\tau > 1.254
    \end{cases}.\label{eq:draine}
\end{multline}
We specifically note that this correction applies to an absorbing foreground slab model. In the equations above, the definitions of all parameters follow equation~\ref{eq:drain-tau}. The parameter $\gamma$ is the damping constant of the Lorentzian profile and is of order 10 for radiative damping. We stress that for \hto\ lines discussed in this paper, the Lorentzian line width that $\gamma$ corresponds to is {10$^{-9}$}~\kms, and is negligible compared to the observed Doppler width (a few~\kms). 

For such an absorbing slab model, the emission from the foreground is negligible against the representative background temperature of $\sim$600~K in this study \citep[see Section~3.2.1 in ][]{li22}. Furthermore, if the foreground cloud does not cover the entire observing beam, a covering factor $f_c$ ($0\leq f_c \leq1$) has to be accounted for and equation~\ref{eq:2} is modified to:
\begin{equation}
    I_v = I_{\textrm{c}} (1 - f_c (1 - \textrm{e}^{-\tau(v)})), \label{eq:ff}
\end{equation}
and the left-hand side of the equation \ref{eq:draine} is modified to $W_\lambda/(\lambda f_c)$.

\subsubsection{Stellar Atmosphere Model of a Circumstellar Disk}\label{subsec:sam}

The absorption can also occur in an accretion disk scenario in the system of a forming massive star \citep{barr20, barr22, li22}. Specifically, the disk has an outward-decreasing temperature gradient from the mid-plane to the surface. Such disks show absorption lines because the thermal continuum from the dust is mixed with the molecular gas. For such a scenario, we adopt the curve-of-growth of the stellar atmosphere model in which the continuum and the line {opacities} are coupled. \textcolor{black}{The residual flux}, 
\begin{equation}
    R_\nu \equiv I_\nu/I_c,
\end{equation}
\textcolor{black}{{{where $I_\nu$ is the intensity of the absorption line,}}} can then be approximated by the Milne-Eddington model \citep[][Ch 10]{mihalas78} which assumes a grey atmosphere. The absorption line profile may originate in pure absorption or scattering. The parameter $\epsilon$ characterizes the line formation, and can take the form of 1 (pure absorption), 0 (pure scattering) or between 0 and 1 (a combination of scattering and absorption). We refer to Appendix~A in \citet{barr20} for details of {the line residual flux expected from} this model. 

The curve of growth in the stellar atmosphere model is constructed via the equivalent width versus $\beta_0$, the ratio of the line opacity at the line center, $\kappa_L(\nu=\nu_0)$, to the continuum opacity $\kappa_c$:
 \begin{equation}\label{eq: mihalas}
  \begin{aligned}
\frac{W_\nu}{2\Delta \nu_D} &= \int_0^{+\infty} (1-R_v) dv \\
     &= A_0\int_0^{+\infty} \beta_0 H(a, v)[1+\beta_0 H(a, v)]^{-1}dv, 
 \end{aligned}
\end{equation}
in which
 \begin{equation}\label{eq:beta0}
  \begin{aligned}
   \beta_0 &= \frac{\kappa_L(\nu=\nu_0)}{\kappa_c}\\ &= \frac{A_{ul}\lambda^3}{8\pi\sqrt{2\pi}\sigma_v}\frac{g_u}{g_l}\frac{N_l}{\sigma_c N_H} \left(1-\frac{g_l}{g_u}\frac{N_u}{N_l}\right),
 \end{aligned}
\end{equation}
{where we have used the continuum opacity, $\kappa_c$, equals $\sigma_c N_H$ where $\sigma_c$ is the dust cross-section per H-atom, $H(a, v)$ is the Voigt function that gives the line profile in velocity space, $v$ is defined as the velocity shift with respect to the line center in units of the Doppler width (see eqn.~\ref{eqn:voigt}).} The damping factor $a=\gamma \lambda/b$ is of {the order of 10$^{-8}$} for H$_2$O ro-vibrational lines. The parameter $A_0$ is the central depth of an opaque line. Its exact value is determined by the radiative transfer model of the surface of the disk and is related to the gradient of the Planck function. For a grey atmosphere and lines in pure absorption, $A_0$ is \mbox{$\sim$ 0.5--0.9} from 900 to 100 K \citep[see Appendix A in][]{barr20}. The dispersion in velocity space, \textcolor{black}{$\sigma_v$}, is transformed from the Doppler parameter, $b/\sqrt{2}$, with which we convert $W_\nu$ to the velocity space:
\begin{equation}
 \frac{W_\nu}{2\Delta \nu_D} = \frac{W_v}{2b}.
\end{equation}
We can disregard the bracketed item in \mbox{equation \ref{eq:beta0}} if stimulated emission is negligible. 

We adopt a value of \mbox{7$\times 10^{-23}$~cm$^2$/H-nucleus} for $\sigma_c$ following \citet{barr20}, as it is appropriate for coagulated interstellar dust \citep{ormel11}. Theoretical fits to these curves of growth will provide abundances relative to the dust opacity. As this value of $\sigma_c$ is adopted in the CO analysis \citep{li22} as well, we are able to derive an absolute water-to-CO ratio once the CO correspondent component to water is identified via the kinematic information (see \S~\ref{subsec:overlap}). 

\section{Results}\label{sec:re}

\begin{deluxetable*}{@{\extracolsep{8pt}}l c cc cc@{}}
\tablecolumns{6}
\tabletypesize{\scriptsize}
\tablecaption{Physical Conditions of Decomposed Components \label{table:sum-prop}}
\tablehead{\colhead{Component} & \colhead{`W', $-45$~\kms}  & \multicolumn{2}{c}{`H1', $-39.5$~\kms} & \multicolumn{2}{c}{`H2', $-54.5$~\kms}   \\
\cline{2-2} \cline{3-4}   \cline{5-6}  
\colhead{Transitions} & \colhead{$\nu_2$ = 1--0} & \colhead{$\nu_2$ = 1--0} & \colhead{$\nu_2$ = 2--1}  & \colhead{$\nu_2$ = 1--0} & \colhead{$\nu_2$ = 2--1} }
\startdata
\multicolumn{6}{c}{Rotation Diagram }\\
\hline
\tex~(K) & 200 $\pm$ 18 & 807 $\pm$ 58 & 703 $\pm$ 60  & $669\pm57$ & $946\pm170$\\
$N_\textrm{tot}$~(\percmsq) & $(6.6\pm4.0)\times10^{17}$\tablenotemark{a} & $(1.2\pm0.6)\times10^{18}$ & $(7.9\pm3.7)\times10^{16}$ & $(6.2\pm3.4)\times10^{17}$ & $(8.1\pm5.6)\times10^{16}$ \\
\hline
\multicolumn{6}{c}{Slab Model}\\
\hline
$f_c$ & ... & 0.4 & 0.4 & 0.3 & ... \\
$b$~(\kms) & ... & 2.8 & 2.8  & 3.5 & ... \\
\tex~(K) & ... & 471$^{+14}_{-15}$ & 654$^{+135}_{-191}$ & 600$^{+28}_{-27}$ & ...  \\
$N_\textrm{tot}$~(\percmsq) & ... & $2.5_{-0.2}^{+0.3}\times 10^{19}$ & $2.8_{-0.7}^{+0.5}\times 10^{17}$  & $5.3_{-0.6}^{+0.3}\times 10^{18}$ & ...  \\
$\chi^2_{r, min}$ & .. &  1.7 & 0.65 & 3.8 & ... \\
\hline
\multicolumn{6}{c}{Disk Model\tablenotemark{b}}\\
\hline 
$b$~(\kms) & ... & 2.8 & 2.8 & 3.5 & ...\\
\tex & ... & 491$_{-14}^{+13}$ & 691$_{-212}^{+122}$ & 612$^{+27}_{-30}$ & ...\\
Abun. (w.r.t H) & ... & $2.6_{-0.2}^{+0.1}\times 10^{-3}$ & $5.0_{-3.5}^{+1.4}\times 10^{-5}$ & $5.1_{-0.5}^{+0.4}\times 10^{-4}$ & ...\\
$\chi^2_{r, min}$ & ... &  1.7 & 0.34 & 2.3 & ...\\
[0.5ex]
\enddata
\tablenotetext{a}{This value is corrected by assuming that `W' covers both proto-binary and the absorption intensity is not diluted (see \S~\ref{subsec:overlap}).}
\tablenotetext{b}{Pure absorption with parameter $\epsilon$ of 1 are assumed in this table. See Table~\ref{table:degeneracy} for results with $\epsilon$ of 0 and 0.5.}
\end{deluxetable*}

As shown in \S~\ref{sec:da}, the H$_2$O $\nu_2$=1--0 lines are decomposed into three velocity components at  $-45, -39.5, $ and $-54.5$~\kms, while the $\nu_2$=2--1 absorption lines are decomposed into two components at $-39.5$ and $-54.5$~\kms. {The components at the three different velocities, as explained in \S~\ref{subsec:bd}, are labeled as `W', `H1', and `H2' based on the preliminary estimation of their temperatures derived from the rotation diagram analysis under the optically thin assumption (Table~\ref{table:sum-prop}).}


We hereafter apply the curve-of-growth analyses to the $\nu_2$=1--0 transitions from `H1 and `H2' and $\nu_2$=2--1 from `H1'. {The rotation diagrams of the three sets of transitions illustrate a large scatter indicative of the opacity problem discussed in \S\ref{subsec:h1h2}, unlike for the other components and transitions.} In \S~\ref{subsec:overlap}, all the water components are compared to and connected with warm and/or hot CO components. The implications of the vibrationally excited water lines are presented in \S~\ref{subsec:vib}.

\begin{figure*}[!t]
    \centering
    \includegraphics[width=0.45\linewidth]{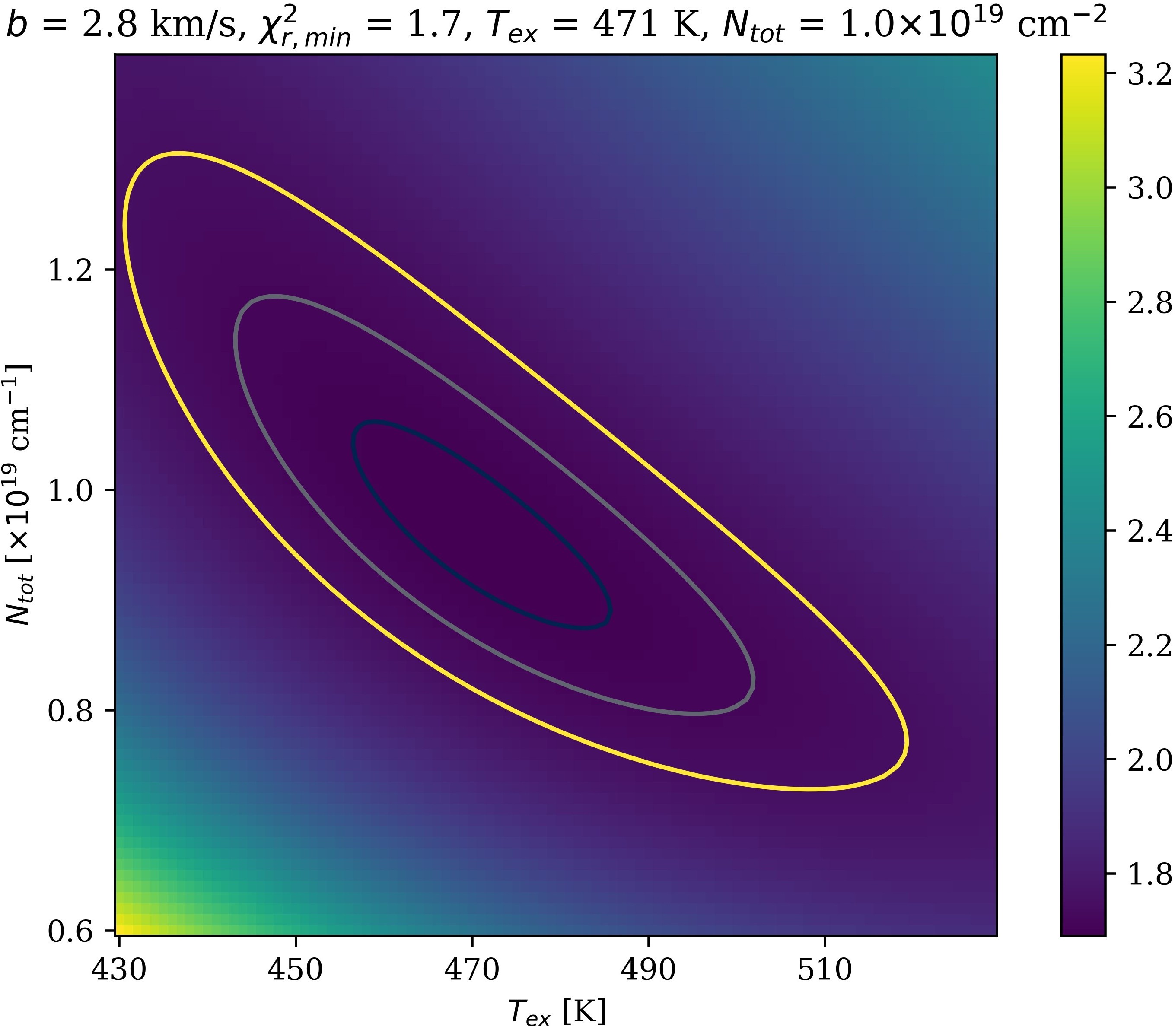}
    \includegraphics[width=0.5\linewidth]{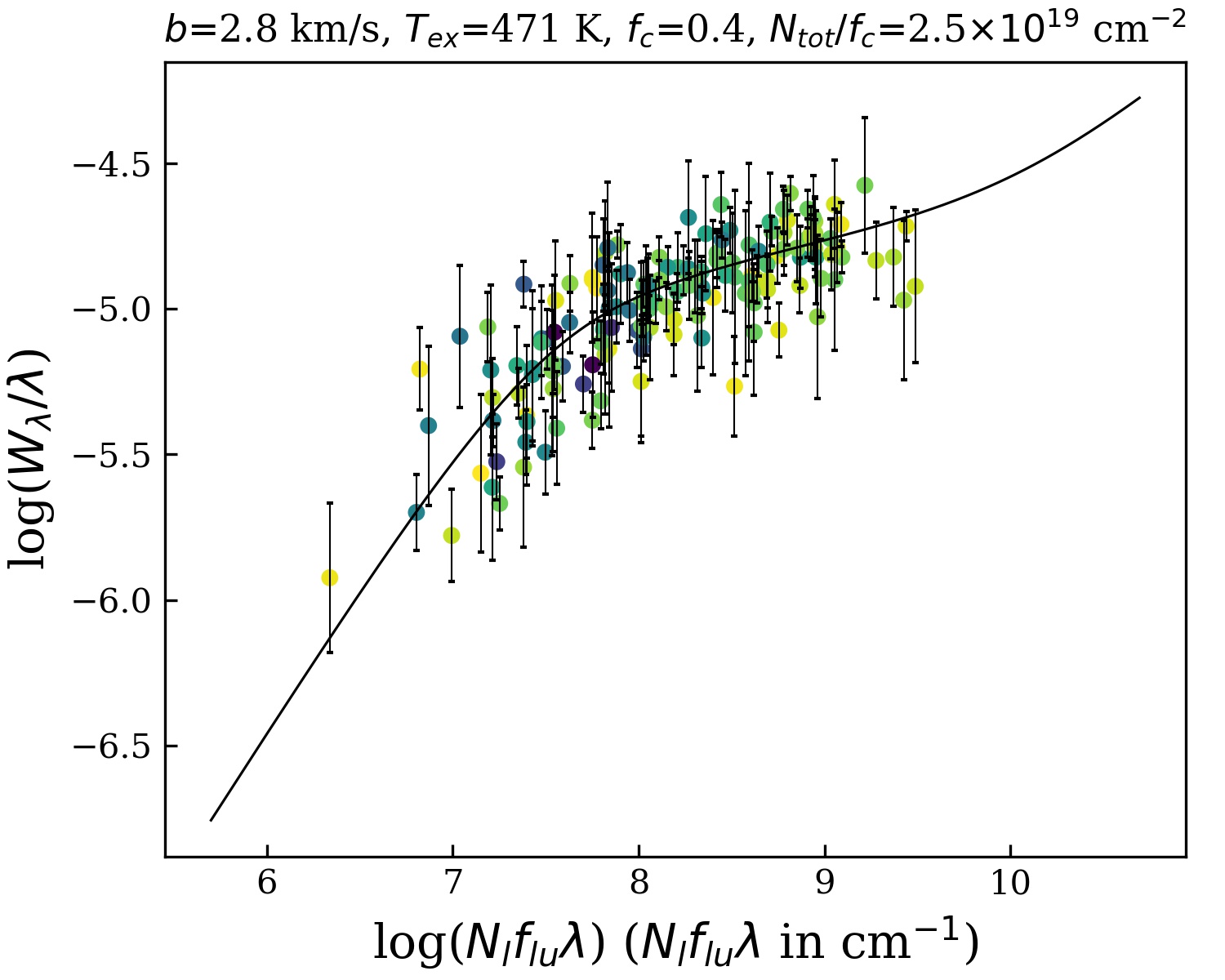}\\
    \includegraphics[width=0.45\linewidth]{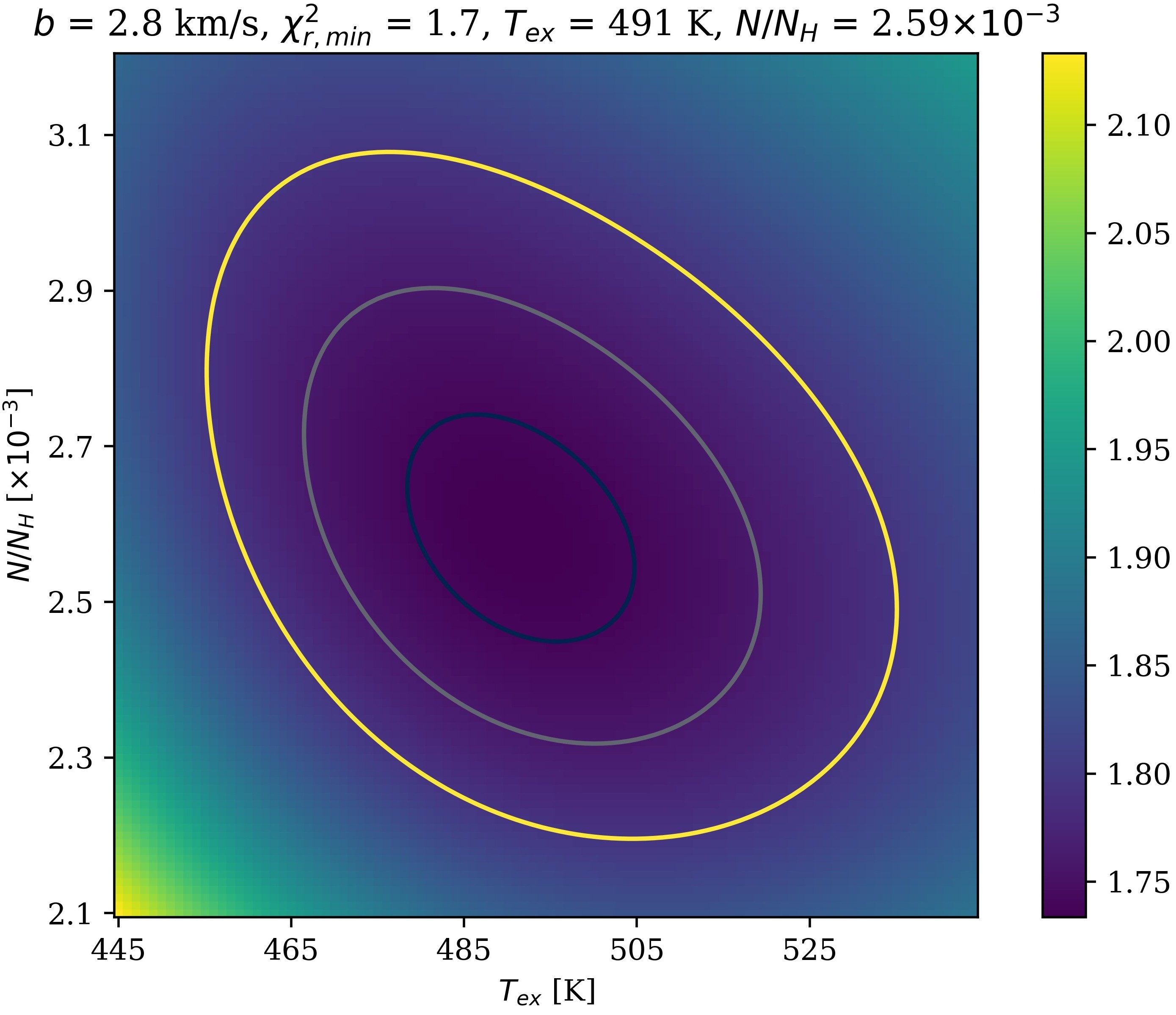}
    \includegraphics[width=0.5\linewidth]{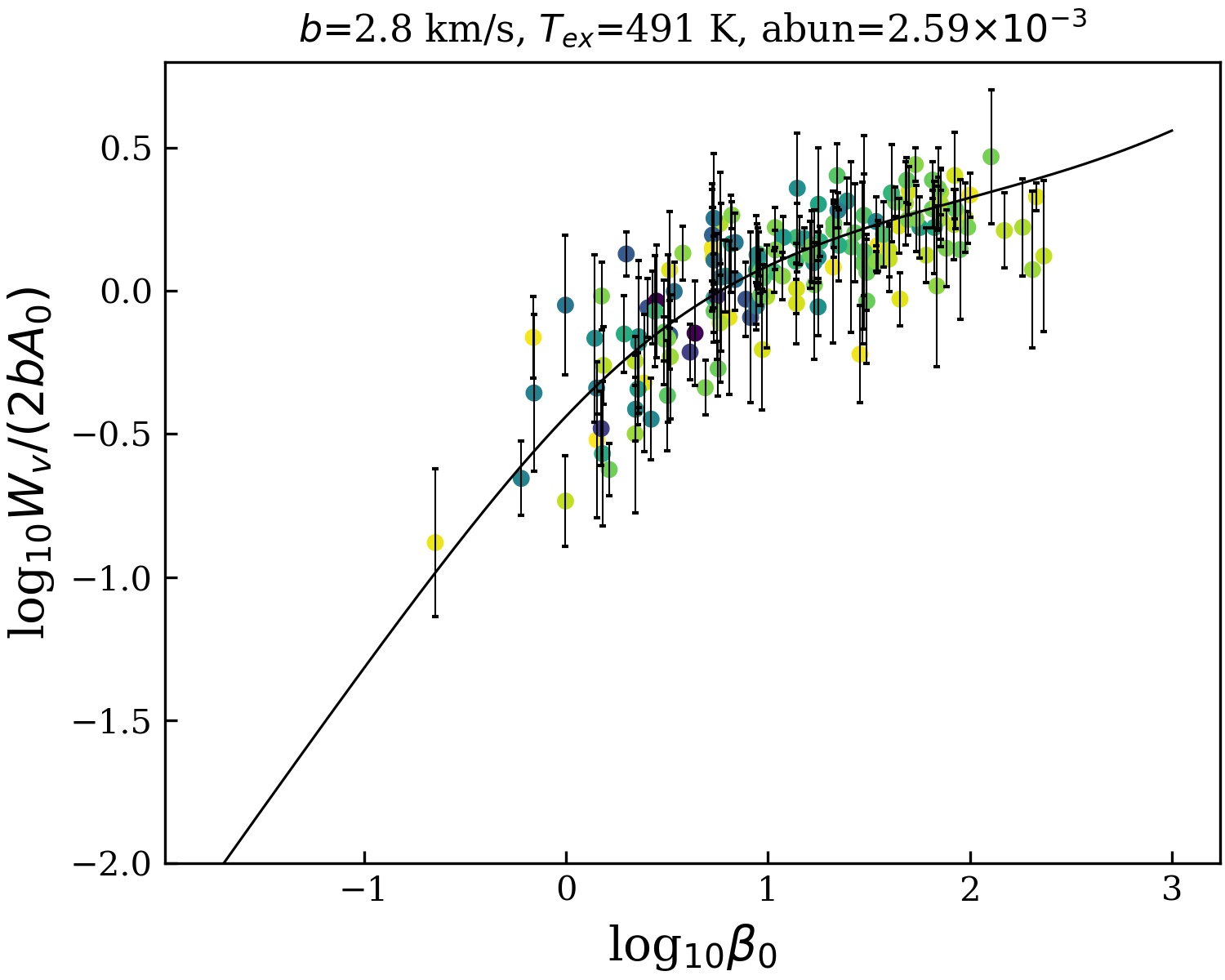}
    \caption{{Curve of growth analysis for the slab (top) and disk (bottom) models for $\nu_2$=1--0 absorption lines from the `H1' component.} \textit{Left panels:} Grid-search results for both the slab and the disk model illustrating the best-fitting results. The contours represent the 1$\sigma$, 2$\sigma$, and 3$\sigma$ uncertainty levels. \textit{Right panels}: the curves of growth for the slab and the disk model. The color scale of the data points is the same as those in the rotation diagram. {Similar figures for the curve-of-growth analysis for $\nu_2$=1--0 transitions from the `H2' component, and $\nu_2$=2--1 transitions from the `H1' component, are presented in Figures~\ref{fig:cog-h2} and \ref{fig:cog-h1-2-1}, respectively.}}
    \label{fig:cog}
\end{figure*}

\begin{deluxetable*}{@{\extracolsep{8pt}}c ccc ccc ccc@{}}
\tablecolumns{10}
\tabletypesize{\scriptsize}
\tablecaption{Results of the physical component `H1' Derived from Different $\epsilon$ and $\sigma_v$ in the Disk Model  \label{table:degeneracy}}
\tablehead{&  \multicolumn{3}{c}{$\epsilon$=0} & \multicolumn{3}{c}{$\epsilon$=0.5} &  \multicolumn{3}{c}{$\epsilon$=1}  \\
\cline{2-4}   \cline{5-7}  \cline{8-10}
\colhead{} & \colhead{$\chi^2_{r, min}$} & \colhead{$T_\textrm{ex}$ (K)} & \colhead{$X_{\textrm{H}_2\textrm{O}}/X_\textrm{CO}$}  & \colhead{$\chi^2_{r, min}$} & \colhead{$T_\textrm{ex}$ (K)} & \colhead{$X_{\textrm{H}_2\textrm{O}}/X_\textrm{CO}$} & \colhead{$\chi^2_{r, min}$} & \colhead{$T_\textrm{ex}$ (K)} & \colhead{$X_{\textrm{H}_2\textrm{O}}/X_\textrm{CO}$} 
}
\startdata
$\sigma_v$=1.5~\kms& 2.02& 579 & 1.19 & 1.79 & 449 & 2.45 & 1.88 & 419 & 3.71 \\ 
$\sigma_v$=2.0~\kms& 2.54 & 565 & 0.78 & 1.84 & 510 & 1.30 & 1.73 & 491 & 1.53 \\ 
$\sigma_v$=2.5~\kms& 3.14 & 598 & 0.61 & 1.99 & 560 & 0.90 & 1.88 & 521 & 1.20 \\ 
$\sigma_v$=3.0~\kms& 3.72 & 622 & 0.52 & 2.37 & 594 & 0.72 & 1.98 & 564 & 0.92 \\ 
\enddata
\tablecomments{(1) The $1\sigma$ uncertainty for the derived temperatures is in order of $\pm10$~K. (2) $X_{\textrm{H}_2\textrm{O}}/X_\textrm{CO}$ is derived by assuming that `H1' and `MIR2-H1' in CO coexist.}
\end{deluxetable*} 

\subsection{Two Hot Physical Components: H1 and H2}\label{subsec:h1h2}

We conduct the \textcolor{black}{grid search method}~\citep{li22} on the (\tex, $N_{\textrm{tot}}$) and (\tex, abundance) grid in the curve-of-growth analyses for the slab model and the disk model, respectively. {For the slab model, \tex\ together with $N_{\textrm{tot}}$ determines $\tau_p$, or $Nf_{ul}\lambda$ ({Equation}~\ref{eq:draine}). We, therefore, search for the parameter combination of (\tex, $N_{\textrm{tot}}$) that gives the smallest reduced $\chi^2$ between the observed equivalent width $W_\lambda$ (in the wavelength space) and that derived from the theoretical curve of growth~({Equation}~\ref{eq:draine}). Similarly, for the disk model, \tex\ together with the abundance ($N/N_\textrm{H}$) determines $\beta_0$, the ratio of the line opacity to the continuum opacity. We search for the best (\tex, abundance) combination that gives the smallest reduced $\chi^2$ between the observed equivalent width $W_\nu$ (in the frequency space) and that derived from the theoretical curve of growth~({Equation}~\ref{eq: mihalas}).    }

When fitting the observed $W_\lambda/\lambda$ to the theoretical curve-of-growth of a slab model, we note that the partial coverage, $f_c$, and the Doppler width, $b$ (=$\sqrt{2}\sigma_v$), are degenerated parameters. Different combinations of $f_c$ and $b$ can provide as good fitting results, and we illustrate this point below for a similar case in the disk model. Observational results can put constraints on the range of $f_c$ and $b$ to some extent. For example, $f_c$ are constrained as $\sim$0.4, which is twice the lower limit of the line intensities, $\sim$0.2. The factor of two is because of the dilution effect, that the absorption line is against the continuum contributed by both MIR sources in \wt\  (see \S~\ref{sec:obs}). As for the Doppler width $b$, its lower limit can be constrained by the thermal line width, and for gas of 500~K, $\sigma_\textrm{th} = \sqrt{(k_\textrm{B}T)/(\mu m_\textrm{H})}$ = 9.12$\times 10^{-2}$ \kms\ ($T$/K)$^{0.5} \mu^{-0.5}$ $\approx0.5$~\kms. The upper limit of $b$ can be constrained by the observed line width, which is $b$ convolved with the instrument resolution $\sigma_\textrm{res}$ = $c/(2\sqrt{2\textrm{ln}2}R)$ = 2.5~\kms. 

For the disk model, there is also a dependence on the chosen line width $b$ and $\epsilon$, the degree of absorption relative to the scattering in the line. We illustrate this in a more quantitative way. Take the `H1' component as an example, as presented in Table~\ref{table:degeneracy}, while the results of the temperature and abundance depend on the adopted $\sigma_v$ and $\epsilon$, combinations of different $\sigma_v$ and $\epsilon$ may provide comparable $\chi^2_{r, min}$. Although we may have some control over $\sigma_v$, the value of $\epsilon$ is unconstrained. We, therefore, provide Table~\ref{table:degeneracy} as a reference for conditions when the lines are not due to pure absorption.




\begin{figure*}[!t]
    \centering
    \includegraphics[width = 0.9\linewidth]{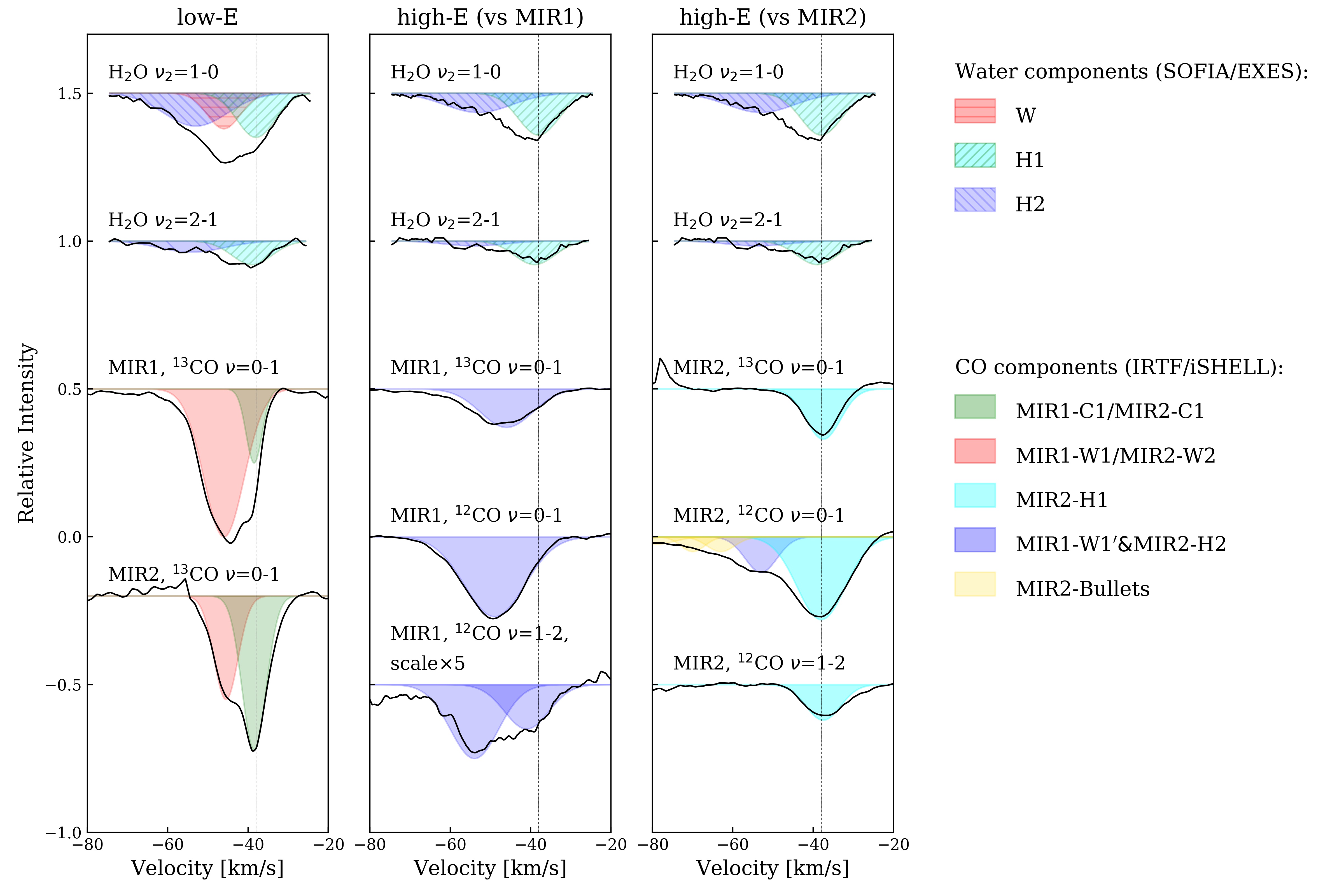}
\caption{Comparison between averaged H$_2$O and CO components in low ($<$400K) and high ($>$400K) energy levels.  The dashed vertical line represents $v_\textrm{sys}=-38$~\kms, and the decomposed components are colored following \citet{li22}. {In principle, components with the same central velocities are given the same color. If components are at the same velocity and are identified with different temperatures, different colors are given. MIR1-W1$^\prime$ is an exception due to the multiple potential origins~\citep{li22}, and for lack of further information both components in the averaged spectra of MIR1 \co~$\nu$=1--2 are given the same color.} }
    \label{fig:overlap}
\end{figure*}

For `H1' and `H2', we therefore choose the partial coverage, $f_c$, to 0.4 and 0.3, and set the Doppler width $b$ (=$\sqrt{2}\sigma_v$) to 2.8 and 3.5~\kms, separately, since The chosen values provides the smallest $\chi^2_{r}$. For the $\nu_2$=1--0 transition, we present the best-fitting results of the `H1' component for both the foreground and disk model in Figure~\ref{fig:cog} (results of `H2' are presented in Figure~\ref{fig:cog-h2} in Appendix~\ref{app:e}) and summarize the derived properties in Table~\ref{table:sum-prop}, assuming that the lines are formed in pure absorption ($\epsilon=0$). In either the slab or the disk model, about half of the data points are located on the logarithmic part of the curve-of-growth, confirming that the corresponding absorption lines are optically thick. In the slab model, the curve-of-growth analysis ``corrects" the underestimated total column densities in the rotation diagram by a factor of 21 and 9 for `H1' and `H2', respectively. The derived temperatures are lowered to 471~K and 600~K, respectively (Table~\ref{table:sum-prop}). In the disk model, as relative abundances are derived, the correction in column densities is quantified once the connection between CO and water components is established (see \S~\ref{subsec:overlap}). The derived temperatures from the disk model are comparable to those derived from the slab model (Table~\ref{table:sum-prop}). As for the $\nu_2$=2--1 transition on `H1', the correction after the curve-of-growth analysis on the temperature and column density is insignificant ({Figure~\ref{fig:cog-h1-2-1} in Appendix~\ref{app:e}} and Table~\ref{table:sum-prop}), indicating that the optical depth effect is not severe.


\subsection{Connecting the Absorbing Components of H$_2$O and CO}\label{subsec:overlap}

\begin{deluxetable}{@{\extracolsep{8pt}}ll cc @{}}
\tablecolumns{4}
\tabletypesize{\scriptsize}
\tablecaption{Comparing  Water and CO Components \label{table:com-co}}
\tablehead{\colhead{Comp.}  & \colhead{Species}   & \colhead{\vLSR}  & \colhead{\tex}   \\\colhead{}  & \colhead{}   & \colhead{(\kms)}  & \colhead{(K)}    }
\startdata
`W' & H$_2$O & $-46$ & 200 $\pm$ 18  \\
MIR1-W1 & CO & $-43$ & 180$^{+11}_{-14}$   \\
MIR2-W2 & CO & $-45.5$ & 116$\pm$7   \\
\hline
`H1' & H$_2$O (Slab) & $-39.4\pm2.1$& 471$^{+14}_{-15}$ \\
& H$_2$O (Disk) & $-39.4\pm2.1$& 491$^{+13}_{-14}$  \\
MIR2-H1 & CO & $-37.5$ & 662$^{+23}_{-28}$ \\
\hline
`H2' & H$_2$O (Slab) & $-54.1\pm3.0$ & 600$^{+28}_{-27}$ \\
& H$_2$O (Disk) & $-54.1\pm3.0$ & 612$^{+27}_{-30}$ \\
MIR2-H2 & CO & $-52$& 482$^{+97}_{-70}$ \\
MIR1-W1$^\prime$ & CO & $-40, \sim-54$& 709$^{+136}_{-101}$\\
[0.5ex]
\enddata
\tablecomments{{Physical conditions of CO components are adopted from Table 7 in \citet{li22}. The adopted temperatures of hot CO components (MIR2-H1, MIR2-H2, MIR1-W1$^\prime$) are corrected values assuming a disk model.}}
\end{deluxetable} 

High spectral resolution $M$-band absorption spectroscopy ($R$=88,100) toward \wt\ at 4.7~\um\ revealed multiple kinematic components in \co\ and its isotopologues~\citep{li22}. Specifically, IRTF/iSHELL spatially resolved the two protostars in \wt. As CO lines also trace ambient gaseous components that absorb against the MIR background, building a connection between the components identified in H$_2$O and CO will help us to understand the origins of the gaseous H$_2$O components.

The velocity resolution of EXES (6~\kms) and iSHELL (3.4~\kms) enables the comparison between H$_2$O and CO components through their kinematic information. {We note that we elected to link the observed H$_2$O and CO components through their similarity in velocity and temperature. The velocity width is not used for aiding the comparison because it is dependent on the optical depth and is degenerated with the column density (Table~\ref{table:degeneracy}).} We, therefore, present in Figure~\ref{fig:overlap} the comparison between the absorption profiles {and list in Table~\ref{table:com-co} the velocity and derived excitation temperature of components identified in the two species}.

As the different velocity components are characterized by different excitation temperatures, both the profiles of the H$_2$O and CO spectra are averaged in low-energy ($<$400~K) and high-energy ($>$400~K) transitions. As we have described in \S~\ref{sec:da}, we identified three components centered at $-39.5, -45,$ and $-54.5$~\kms\ {(`H1', `W', and `H2')} in low-energy transitions, and two centered at $-39.5$ and $-54.5$~\kms\ in high-energy transitions. We note that half of the transitions (6 of 12) with a $-45$ km/s component are at energies between 400 and 800~K. 

In low-energy levels, both MIR1 and MIR2 have two components revealed in CO, with one centered at the $v_\textrm{sys}=-38$~\kms~{(MIR1-C1 and MIR2-C1\footnote{Naming of each individual CO components is consistent with the nomenclature in \citet{li22}. More specifically, each CO component has a prefix of either ``MIR1-" or ``MIR2-".})}, and the other centered at $-46$~\kms~{(MIR1-W1 and MIR2-W2)}. No low-energy CO components were found at $-55$~\kms. MIR1-C1 and MIR2-C1, so as MIR1-W1 and MIR2-W2 have similar temperatures and column densities~\citep[see Table 7 in][so as temperatures of CO components mentioned below]{li22}, and are regarded to cover MIR1 and MIR2 simultaneously. 

{The temperatures of MIR1-C1 and MIR2-C1 are low ($\sim$50~K). MIR1-C1 and MIR2-C1 are therefore not related to any water components. In contrast, the temperatures of MIR1-W1 and MIR2-W2 are $\sim$180~K and are close to that of the H$_2$O `W' component (\tex$\sim$200~K, Table~\ref{table:sum-prop}) which is at $-46$~\kms, and we thus conclude that MIR1-W1/MIR2-W2 corresponds to the H$_2$O `W' component. We emphasize that, because the CO components MIR1-W1 and MIR2-W2 were found to be in front of both binary stars, for the `W' component in H$_2$O, we do not consider the dilution of the relative intensity (see discussion in \S~\ref{sec:obs}) when performing the rotation diagram analysis in \S~\ref{sec:da}.
}






{We note that for the `H1' H$_2$O component which is centered at $-39.5$~\kms\ and is close to $v_\textrm{sys}$, the existence of $\nu_2$=2--1 transitions and its derived high temperature exclude a connection between it and the cold CO component (`MIR1-C/MIR2-C'). It is possible that in the low energy CO transitions, the CO counterpart of this hot H$_2$O component at $-38$~\kms\ may be hidden underneath a lower temperature component at that velocity. We will discuss this possibility in \S~\ref{subsubsection:foreground}.}

At high-energy levels, MIR1 and MIR2 contribute differently toward the absorbing components. The component MIR2-H1 in CO is centered at $-37.5$~\kms\ and is characterized by 600--700~K. We link this to the `H1' water component. It is more difficult to determine the origin of the $-54.5$~\kms\ `H2' H$_2$O component. {On the one hand, the velocity position, as well as the temperature, of the `H2' H$_2$O component, are compatible with the CO MIR2-H2 component (Table~\ref{table:com-co}). On the other hand, the `H2' H$_2$O component is also likely correlated with the hot CO MIR1-W$^\prime$ component, which is a complex amalgam of several components. MIR1-W$^\prime$ has a complicated origin as is indicated by the varying average line profiles in \thco, \co, and \co\ \nuu=2--1. One of the two peaks in the \co\ \nuu=2--1 profile (at $-54$~\kms) of MIR1-W$^\prime$ coincides more or less with the $-54.5$~\kms\ H$_2$O `H2' component, although the other one (at $-39.5$~\kms) has no counterpart in the H$_2$O spectrum. Likewise the \co\ and \thco\ MIR1-W$^\prime$ components are centered at $-46$ and $-49$~\kms, respectively. Hence, we consider that the H$_2$O `H2' component is related to one of the CO MIR1-W$^\prime$ components.} 


In conclusion, the water component `H1' is much stronger and it better matches MIR2-H1 in CO. The water component `H2' is weaker and its origin is less clear. {Thus we connect `W' to the warm component at $-45$~\kms\ in CO, `H1' to MIR2-H1, and `H2' to either MIR2-H2 or MIR1-W$^\prime$. Once the hot components in H$_2$O are linked to those in CO, we can derive the H$_2$O/CO abundance ratio (Table~\ref{table:all}).} 
{Assuming that the CO abundance is equal to the gas phase C abundance in the diffuse ISM~\citep[1.6$\times 10^{-4}$,][]{cardelli96, sofia97},} we derive H$_2$O column densities under the disk model of 3.6$\times10^{19}$ and 8.9$\times10^{18}$~\percmsq\ for `H1' and `H2', respectively (Table~\ref{table:all}).


\section{Discussion}\label{sec:dis}

\begin{deluxetable*}{lccccc}[!t]
\tablecolumns{6}
\tabletypesize{\scriptsize}
\tablecaption{Comparison of H$_2$O characteristics Derived from ISO-SWS and SOFIA-EXES Observations \label{table:all}}
\tablehead{ \colhead{Properties} & \colhead{ISO-SWS\tablenotemark{a}} & \multicolumn{4}{c}{SOFIA-EXES} \\
\cline{3-6}
\colhead{} & \colhead{} & \colhead{H1 (Slab)} & \colhead{H1 (Disk)} & \colhead{H2 (Slab)} & \colhead{H2 (Disk)}}
\startdata
$T_\textrm{ex}$(H$_2$O) (K) & 400$^{+200}_{-150}$ & 471$^{+14}_{-15}$ & 491$_{-14}^{+13}$ & 600$^{+28}_{-27}$ & 612$^{+27}_{-30}$ \\
$N$(H$_2$O) (\percmsq) & 3$_{-1}^{+1}\times10^{17}$ & 2.5$_{-0.2}^{+0.3}\times10^{19}$ & 3.6$\pm 1.2 \times10^{19}$ & 5.3$^{+0.3}_{-0.6}\times10^{18}$ & 8.9$\pm 0.3 \times10^{18}$\\
$X$[H$_2$O]/$X$[CO] & 0.05 & 1.1$^{+0.4}_{-0.4}$ &  1.5$\pm 0.5$ & 3.8$^{+1.8}_{-2.0}$ \textit{or} 0.9$^{+0.2}_{-0.4}$ & 0.9$\pm$0.3 \\
\enddata
\tablenotetext{a}{\citet{boonman03}.}
\end{deluxetable*} 

\begin{deluxetable*}{lc llr}[!t]
\tablecolumns{5}
\tabletypesize{\scriptsize}
\tablecaption{{Identified CO Components from the Perspective of H$_2$O }\label{table:co-water}}
\tablehead{ \colhead{CO} & \colhead{H$_2$O} & \colhead{Origins} & \colhead{Implications} & \colhead{Refs.}}
\startdata
MIR2-H1 & `H1'& Blobs on the disks & The inclination of disks are in question& \ref{subsec:po-hot}\\
MIR1-W1$^\prime$/MIR2-H2 & `H2'& Blobs on the disks & The inclination of disks are in question& \ref{subsec:po-hot} \\
MIR1-W1/MIR2-W2 & `W' & Shared warm foreground component& Low H$_2$O/CO relative abundance & \ref{subsubsection:foreground}\\
MIR1-C1/MIR2-C2 & Not detected & Shared cold foreground {envelope} & Low H$_2$O/CO relative abundance& \ref{subsubsection:foreground}\\
MIR2-B1 to B4 & Not detected & Shocked bullets& Maybe not linked to water masers& \ref{subsec:bullets} \\
\enddata
\end{deluxetable*} 

Under the framework of the known structures in \wt, the $M$-band spectroscopic study on CO pictures the kinematic and physical properties of the gaseous environment. Specifically, it includes a shared foreground envelope at $-38$~\kms, several high-velocity clumps (referred to as ``bullets") from $-60$ to $-100$~\kms, and a few warm and/or hot components in the immediate environment of the binary from $-38$ to $-60$~\kms~\citep{li22}. While we have connected `W', `H1', and `H2' in water to MIR1-W1/MIR2-W1, MIR2-H1, and MIR2-H2 (or MIR1-W$^\prime$) in CO (\S~\ref{subsec:overlap}), gaseous water is not detected in the shared cold envelope or the bullets. We refer to past water studies for a more complete view toward \wt~(see Table~\ref{table:all}) in this section and discuss the implications of these ``detections" and ``non-detections" from the perspective of both the kinematics and the chemical abundances~(Table~\ref{table:co-water}).

\subsection{Hot Gaseous Water in \wt}\label{subsec:hot} 

\subsubsection{Physical Origins of the Hot Components}\label{subsec:po-hot}

As we have built the connections of H$_2$O `H1' and `H2' components to the hot CO components via the kinematic information, it is natural to consider that, as for the corresponding CO components, the two H$_2$O components have a disk rather than a foreground cloud origin. Although the observed $W_\lambda/\lambda$ can be successfully fitted to the theoretical curve-of-growth of either model (see \S~\ref{sec:da} and \S~\ref{subsec:h1h2}), and the two models derive comparable column densities (see \S~\ref{subsec:overlap}) and temperatures, we nevertheless consider the disk model as the more preferred one for the following reasons.

If the hot components have a foreground origin, the heating mechanisms are either due to the radiative heating or due to shocks. On the one hand, if the water components are radiatively heated, the temperature range from 450--600~K corresponds to a distance of 280--140~au in the \wt\ system \citep{li22}. As `H1' is at the systemic velocity, this would imply that this component is static in radial velocity in the highly dynamic environment close to a high mass protostar. The `H2' component has the opposite issue as it is moving at a radial velocity of $-15$~\kms\ relative to the system, and would move outward by some 100~au in 30 years. As the physical conditions derived from CO observations in 2020 are very similar to those derived in 1991~\citep{mitchell91}, this distance range does not work for H2, either, because of its relative velocity of 15~\kms, and it shall move outward along the line of sight for a large distance ($\sim$100~au). On the other hand, interpreting the warm temperature of these two components as the results of shocks has issues too. For J-shocks, the derived temperatures are quite high. J-shocks initially heat the gas to very high temperatures larger than 10$^4$~K. The gas then cools rapidly and, as molecules form, the H$_2$ formation energy keeps the temperature at $\sim 400$ K for a column density of $\sim 10^{22}$ cm$^{-2}$~\citep{hem13}. Using the relationship between postshock temperature in this plateau region derived from J shock models~\citep[eq. 24 in][]{hem13}, a temperature of 500 K (c.f., Table~\ref{table:all}) requires a preshock density of, $n_0\simeq 10^8\, \left(T/500\, \text{K}\right)^{8.3} \, \left(v_s/50\, \text{km/s}\right)\, \left(\Delta v_D/1\, \text{km/s}\right)^{1.8}$ cm$^{-3}$, where $v_s$ is the shock velocity and $\Delta v_D$ the Doppler width in the molecular gas. C-type shocks can yield temperatures of 450--600~K for shock velocity of $\sim$10~\kms~\citep{kn96}. However, a C-shock produces a warm H$_2$ column density of $\sim10^{21}$~\percmsq, which is far too small to be consistent with the column density we derived for the hot components.

If `H1' and `H2' are in the disks, the H$_2$O (and CO) excitation temperatures are similar{ to }(but slightly lower than) the dust continuum temperature. Hence, heating is not an issue. However, the problem now resides in the exact positions of the two components on the disk based on their projected velocity information along the line of sight. As pointed out in \citet{li22}, it is difficult to pinpoint the locations of the blobs on the disk because the inclination angles of both MIR1 and MIR2 and the systematic velocity of MIR1 are unknown. {`H1' was connected with `MIR2-H1' because of the similar velocity and temperature. We note that although the central velocities of `H1' and `MIR2-H1' differ by $\sim$2~\kms, such a difference is insignificant compared with the uncertainty level of {\vLSR} of `H1', which is 2.1~\kms~(see \S~\ref{sec:da}).} 
For `H2', it is not even clear whether it is associated with MIR1 or MIR2, as the binary protostars are not spatially resolved by SOFIA, and both binary stars show hot components at the same velocity close to $-55$~\kms. As shown in Figure~\ref{fig:overlap}, in MIR1, the hot component shows up in \co\ \nuu=2--1 vibrationally excited transitions which indicate a high-density region, $\sim10^{10}$~\percmcu. In MIR2, the component is possibly a blob on an inclined disk at a distance smaller than 80~au to the central protostar, but such a scenario poses questions on the inclination of the disk MIR2 again. Therefore, we emphasize that disks in MIR1 and MIR2 need to be spatially and spectrally resolved to fully understand the structures in this region. Observations of vibrationally excited lines via submillimeter interferometers may be very instrumental in settling these issues. 

Analysis of AFGL~2136 and AFGL~2591 observations have faced the same issues in determining whether the hot 600~K absorbing components reside in the foreground slab or the disk~\citep{barr22}. A foreground slab model places very strong constraints on the geometry. In AFGL~2136, if the slab has a water maser origin, the spatial coverage is even larger than the MIR disk, inconsistent with the required covering factor in order to explain the saturation of absorption lines at non-zero flux. In AFGL~2591, wavelength-dependent covering factors are needed to interpret the difference of the spatial coverage derived from the 7 and 13~\um\ spectroscopy, while the component does not cover the source at all at 3~\um~\citep{barr22b}. Continuum emission size and chemistry need to be radius dependent, possibly due to the temperature gradient, to interpret the different covering factors. Moreover, similar to \wt, too high a density, and too high an abundance argue against a shock origin. In contrast, Keplerian disks as well as clumpy substructures were spatially resolved by Atacama Large Millimeter/submillimeter Array \citep[AFGL~2136; ][]{maud19} and NOEMA \citep[AFGL~2591; ][]{suri21}, supporting the scenario that the absorption arises in blobs in the disk.

In summary, from a broad view, it is a prevalent scenario that hot absorption gas is detected against the MIR continuum backgrounds in massive protostars \citep[e.g.][]{vh96, cern97, lahuis00, boonman03}. Locating a blob on a disk that has a vertical outward-decreasing temperature gradient requires fewer constraints on the geometry than a foreground slab model does. However, disk models face challenges in realizing such an internal heating mechanism. As discussed in \citet{barr22},  the flashlight effect may ensure the disk is not externally heated \citep{nakano89, yorke99, kuiper10}. However, if one proposes the dissipation of gravitational energy, the accretion rate would be orders of magnitude higher than the expected accretion rate~\citep{mc03, hoso10, kuiper11, cog17}. Hence dissipation of turbulent and/or magnetic energy inherited from the prestellar core would be required, implying a very early and active stage in the formation of these massive protostars. 

\subsubsection{{Vibrationally Excited H$_2$O}}\label{subsec:vib}

We presented in Figure~\ref{fig:bd} the rotational temperatures of `H1' and `H2' in the first excited vibrational state ($\nu_2=1$), which are \textcolor{black}{703}$\pm$60~K and 946$\pm$170~K, respectively. Applying curve-of-growth analyses to `H1', the corrected rotational excitation temperatures are 654$^{+135}_{-191}$~K in the slab model and 691$^{+122}_{-212}$~K in the disk model (Table~\ref{table:sum-prop}) and are not far away from the result derived from the rotation diagram. The increment of the total column density of `H1' after correction is $\sim$3--4 times, much less than that of $\sim$20 times on the $\nu_2=0$ state, indicating a less severe optical depth effect.

One can derive the vibrational excitation temperature, $T_\textrm{vib}$ via the Boltzmann equation by comparing the column density in the $\nu_2=0$ level, $N_0$, with $N_1$ in the $\nu_2=1$ level:
\begin{equation}
N_1/N_0  = \textrm{exp}(-2294.7~\textrm{K} /T_{\textrm{vib}}).  \label{eq:tvib}
\end{equation}
We list in Table~\ref{table:vib} multiple vibrational excitation temperatures for `H1' and `H2', using $N_0$ and $N_1$ before and after corrections on the optical depth effects. Results in Table~\ref{table:vib} indicate that such a correction also decreases the derived vibrational excitation temperatures by a few hundred Kelvin. Comparing the corrected $T_\textrm{vib}$ with the corrected rotational excitation temperature for the population in the $\nu_2=0$ state (Table~\ref{table:sum-prop}) or in the $\nu_2=1$ state, we conclude that those temperatures are in relatively good agreement within the error and that vibrational equilibrium is reached.

\begin{deluxetable}{@{\extracolsep{35pt}}l rr @{}}[!t]
\tablecolumns{3}
\tabletypesize{\scriptsize}
\tablecaption{{Vibrational Excitation Temperatures} \label{table:vib}}
\tablehead{\colhead{}  & \colhead{H1}   & \colhead{H2} }
\startdata
$T_\textrm{vib}$(thin) (K) \tablenotemark{a} & 843$^{+553}_{-233}$ & 1127$^{+2083}_{-498}$ \\
$T_\textrm{vib}$(slab) (K) \tablenotemark{b}&  511$^{+30}_{-42}$ & 549$^{+86}_{-130}$ \\ 
$T_\textrm{vib}$(disk) (K) \tablenotemark{c} & 581$^{+52}_{-139}$ & 488$^{+66}_{-100}$ \\
\enddata
\tablenotetext{a}{Values of $N_1$ and $N_0$ in equation~\ref{eq:tvib} are derived from rotation diagram analysis (see Table~\ref{table:sum-prop}; same for the columns below). }
\tablenotetext{b}{$N_1$ and $N_0$ are corrected under the slab model.  Since no curve-of-growth analysis is applied for the $\nu_2$=1 state of `H2', we adopt $N_1$ from the rotation diagram analysis. Same for the `H2' result under the disk model.}
\tablenotetext{c}{$N_1$ and $N_0$ are corrected under the disk model.}
\end{deluxetable}

The existence of vibrationally excited H$_2$O implies that the physical conditions of the hot absorbing gas are extreme as the $\nu_2=1$ state lies 2295~K above the ground vibrational state. The two main excitation mechanisms are collisional excitation due to warm, dense gas and radiative excitation by infrared radiation due to warm dust. If the $\nu_2=1$ state is collisionally populated, a (postshock) density exceeding 10$^{10}$~\percmcu\ is required for thermalization. This order takes account of a critical density\footnote{Take the transition $\nu_2=1$, $J_{K_a, K_c}=1_{0,1}$ as an example, the critical densities from 200 to 1000~K are $\sim6\times10^{10}$ to 10$^{11}$~\percmcu. Values of the Einstein~$A$ and the collisional rate of relevant energy states are adopted from \citet{tennyson01} and \citet{faure08}.}  of 10$^{11}$~\percmcu\ and a radiative trapping effect with $\beta$ of $\sim0.1$ for an optically thick line ($\tau\sim10$). As a comparison, \citet{barr20} estimated a blob density of 10$^9$~\percmcu\ in the disk systems of AFGL~2136 and AFGL~2591.

Other than the collisional excitation, one can estimate the relative importance of the excitation due to the strong radiation field. As is described in \citet[Ch 2.3.3,][]{tielens05}, with a dilution factor $W$ ($W<$1) on the radiation, 

\begin{equation}
    \frac{T_\textrm{ex}}{T_R} = 1 + \frac{kT_\textrm{ex}}{h\nu}\textrm{ln}W,
\end{equation}
in which $T_R$ is the radiation temperature and can be characterized by a dust temperature $T_d$ inside a blackbody. Taking that the hot components are on the surface of the disk and are receiving radiation with $W=0.5$ (half of the disk), that $T_\textrm{ex}$ is from 400--500~K (equation~\ref{eq:tvib}), and that $h\nu$ are larger than 2295~K, we derive ${T_\textrm{ex}}/{T_R}>0.85$. This result indicates that the radiation field does drive the gas to the radiative temperature. We note that at the high implied densities ($\sim10^{10}$~\percmcu), collisions between gas and dust will lead to gas kinetic temperatures that are coupled to but slightly lower than the dust temperature \citep{takahashi83}.

\citet{li22} detected vibrationally excited CO transitions and derived a rotational excitation temperature, 791~K, of CO $\nu=1$ state for the MIR2-H1 component. The similarity excitation temperatures of the vibrational band of water (703~K) and CO support that the excitation temperature is more representative of the color temperature of the radiation field than the kinetic temperature because water has a much larger dipole moment than CO and therefore much more rapid spontaneous radiation. Finally, we emphasize that the scenario that the kinetic temperature is less than the dust temperature is inherent to our results as otherwise, one would observe emission rather than absorption lines against the MIR continuum of the observed sources \citep[see Appendix A in][]{barr22}.

\subsubsection{Comparison with Past Observations}

The temperature range of `H1' and `H2' at $-39.5$ and $-54.5$~\kms\ from 400--500~K is comparable with that derived from the ISO-SWS observations in \citet{boonman03}\footnote{{According to \citet{salgado12}, in the mid-IR, W3~IRS~5 is isolated, even in the large $\sim$20$''$ ISO/SWS beam. To our knowledge, there is only one other source, which is found and labeled as ``MIR3" by \citet{vdt05}. The brightness of MIR3 is $\sim1\%$ of either MIR1 or MIR2, so we ignore its influence on our comparison between the ISO and SOFIA observations.}}.  In the ISO-SWS study, the individual velocity components were not spectrally resolved, separate transitions were blended ($R=1400$, 214~\kms), and a Doppler width $b$ of 5~\kms\ was assumed in modeling the absorption features. As presented in Table~\ref{table:all}, the column density of hot gaseous components derived from our SOFIA/EXES study is about two orders of magnitude larger than the ISO-SWS results. This is a significant increment and much more that the 2.4 and 4.3 times increments derived for AFGL~2136 and AFGL~2591~\citep{barr22}. We interpret this from two aspects: firstly, the absorption intensities of \wt\ measured by ISO-SWS are much lower than those for AFGL~2136 and AFGL~2591~\citep{boonman03}. This indicates a more significant opacity effect than in AFGL~2136 and AFGL~2591. As a result, the column densities in \wt\ corrected by the curve-of-growth analysis are much higher than those corrected values in AFGL~2136 and AFGL~2591 \citep{barr22}. Secondly, saturated lines do not go to 0 but rather reach a non-zero intensity because of either the temperature gradient in a disk atmosphere or a covering factor less than 1 for a foreground cloud. 

\subsection{Other Foreground Gaseous Components}

\subsubsection{The Radiatively Heated Foreground Clouds}\label{subsubsection:foreground}

Comparison between the average low-energy CO and H$_2$O lines (Figure~\ref{fig:overlap}) reveals a warm component `W' at $-45.5$~\kms\ which has a H$_2$O-to-CO relative abundance of 4.4\%. While `W' is considered a shared component in front of MIR1 and MIR2, according to \citet{li22}, this component is radiatively heated and located at least as close as 2000~au to the protostars. In contrast, the cold CO component at $-38$~\kms, which is regarded as a shared foreground envelope of $\sim$50~K, is not present in H$_2$O. If the water in the cold envelope has a comparable column density to that of CO, saturated absorption lines will be detected (see Appendix~\ref{section:hidden}). We conclude that the non-detection of water is due to a too-low column density ($<4.7\times10^{15}$~\percmsq). Therefore, both the warm and the cold components have a low H$_2$O/CO relative abundance.

However, past observations of water lines reveal the rather cool ($\sim$50~K) component but did not observe the warm component ($\sim$200~K). Observations in the pure rotational ortho-lines of water by SWAS \citep{snell00} and Odin \citep{wilson03} derive comparable results. Adopting a temperature of 40~K, both studies reveal a relative ortho-H$_2$O abundance of order 1--2$\times 10^{-9}$, or column densities of $\sim10^{13}$~\percmsq. These results are also comparable with the column density of $10^{13}$~\percmsq\ derived by \textit{Herschel}-HIFI observations~\citep{char10}, albeit that the latter result is rather model dependent. Therefore, we are reporting an EXES upper limit that is much higher than the column density observed by SWAS, Odin, and \textit{Herschel}. We suggest that both the SWAS and Odin beams are very large and they may be measuring the large-scale core, which has a low average column density. If there is a density gradient rising toward the central source, then the submillimeter observations would measure column density that could be much less than along a pencil beam. With a pencil beam, it is difficult to come up with a clear picture of the structural relationship of these three components and with the larger scale structure of the source. 

\subsubsection{The Foreground Bullets}\label{subsec:bullets}

The four high-velocity ``bullets" from $-100$ to $-60$~\kms\ in CO (200--300~K) were not detected in the water observations. Those bullets have been attributed to shocked gas intercepted by the pencil beam~\citep{li22}. Specifically, \citet{li22} quantified the column density, the density, the velocity, and the thickness of these bullets and concluded that they are possibly correlated with the maser clumps moving toward us. As a comparison, assuming that water has a comparable column density to CO, one would expect to see water absorption lines with a depth of $\sim$80$\%$ relative to the continuum. However, among all the identified water lines, the only transition that has a potential absorption feature with an intensity depth of 5$\%$ at $\sim-80$~\kms\ is $2_{2,1}-3_{1,2}$, which has an expected line depth of $\sim30\%$. Therefore, CO bullets are indeed not detected in water lines.

J and C shocks are expected to lead to high abundances of H$_2$O, comparable to CO~\citep{h09, hem13}. Hence, the absence of water absorption lines associated with this high-velocity gas sheds some doubt on their interpretation as shocked bullets and a potential link to water masers. {We emphasize that we do recognize other factors that are related to non-detections of the bullets, although those factors are insignificant.} For example, we may define the baseline beyond $\sim-75$~\kms\ poorly where the bullets are expected. In addition, those foreground bullets were exclusively found in front of MIR2 in the CO observations and may suffer from extra dilution in the SOFIA observations. However, these factors are insignificant for the high abundances of H$_2$O-to-CO, as one would expect to observe prominent saturated H$_2$O absorption features.

\subsection{Chemical Abundances along the Line of Sight}

\begin{deluxetable*}{@{\extracolsep{8pt}}lrrrrrrrrr@{}}[!t]
\tablecolumns{10}
\tabletypesize{\scriptsize}
\tablecaption{The Repository of Elemental Carbon and Oxygen in the Cold Regions of Massive Protostars\label{table:abun}}
\tablehead{&  \multicolumn{3}{c}{\wt} & \multicolumn{3}{c}{AFGL~2136} &  \multicolumn{3}{c}{AFGL~2591}  \\
\cline{2-4}  \cline{5-7} \cline{8-10}
& \colhead{$N$} & \colhead{$X_\textrm{C}$}  & \colhead{$X_\textrm{O}$} & \colhead{$N$} & \colhead{$X_\textrm{C}$}  & \colhead{$X_\textrm{O}$}& \colhead{$N$} & \colhead{$X_\textrm{C}$}  & \colhead{$X_\textrm{O}$}\\
& \colhead{(\percmsq)} & (ppm) & (ppm) & \colhead{(\percmsq)} & (ppm) & (ppm) & \colhead{(\percmsq)} &  (ppm) & (ppm) }
\startdata
Hydrogen & 2.0(23) &  &  & 7.4(22) &  &  & 8.0(22) &  &  \\
\hline
CO (gas)\tablenotemark{a} & 4.7(18) & 23.5 & 23.5 & 5.1(18) & 68.9 & 68.9 & 2.5(18) & 31.3 & 31.3\\
H$_2$O (gas)\tablenotemark{b} & $<$4.7(15) & ... & $<$0.02 & ... & ... & ... & ... & ... & ... \\
CO (ice)\tablenotemark{c} & 2.1(17)& 1.1 &  1.1 & 2.7(17) & 3.6 & 3.6 & ... & ... & ... \\
CO$_2$ (ice)\tablenotemark{c} & 7.1(17) & 3.6 & 7.1 & 7.8(17) & 10.5 & 21.1 & 1.6(17) & 2.0 & 4.0 \\
H$_2$O (ice)\tablenotemark{c} & 5.1(18)& ... & 25.5 & 5.1(18) & ... & 68.9 & 1.2(18) & ... & 15.0 \\
CH$_3$OH (ice)\tablenotemark{c} & 1.7(17)& 0.9 & 0.9 & 2.6(17) & 3.5 & 3.5 & 1.7(17) & 2.1 & 2.1 \\
\hline
Sum (ice) & ... & 5.6 &  34.6 & ... & 17.6 & 97.1 & ... & 4.1 & 21.1 \\
Sum (total) & ... & 29.1 & 58.1 & ... & 86.5 & 166.0 & ... & 35.4 & 52.4 \\
\hline
Diffuse Clouds (gas)& & 160\tablenotemark{d} & 284\tablenotemark{e} & & 160 & 284& & 160 & 284 \\
Silicate Abundance\tablenotemark{f}& & & 200& & & 200& & & 200 \\
Solar Abundance\tablenotemark{g} & &  269& 490& &  269 & 490& & 269 & 490 \\
B-Stars Abundance\tablenotemark{h} & &  214& 575& &  214 & 575& & 214 & 575 \\
[0.5ex]
\enddata
\tablenotetext{a}{Results of gaseous CO are adopted from \citet{li22} for W3~IRS~5, and are from \citet{barr20} for AFGL~2136 and AFGL~2591. The temperatures of the cold regions are $\sim$50~K, 27~K, and 49~K for \wt, AFGL~2136, AFGL~2591, separately \citep{li22, barr20}.}
\tablenotetext{b}{Results from this work.}
\tablenotetext{c}{All measurements on ice are adopted from \citet{gibb04}.}
\tablenotetext{d}{\citet{cardelli96, sofia97}.}
\tablenotetext{e}{\citet{cartledge04}.}
\tablenotetext{f}{\citet{tielens87}.}
\tablenotetext{g}{log$\epsilon_\textrm{C}$ = 8.43, log$\epsilon_\textrm{O}$ = 8.69 \citep{asplund09}.}
\tablenotetext{h}{\citet{np12}.}
\tablecomments{(1) For column densities, powers of 10 are given in parentheses. (2) $X_\textrm{C}$ and $X_\textrm{O}$ are the relative abundances derived from $N_\textrm{C}$/$N_\textrm{H}$ and $N_\textrm{O}$/$N_\textrm{H}$.} 
\end{deluxetable*}

The availability of data on column densities of different species such as gaseous CO and ices along the line of sight toward \wt\ makes it an appropriate example to address the oxygen and carbon budget. We discuss below the reservoirs of the two elements in different environments of the \wt\ system and other massive protostars including the hot disks as well as the cold foreground clouds.

As described in \S~\ref{subsec:hot}, SOFIA observations derived much higher column densities of hot gaseous water {than that of hot gaseous CO} in \wt\ as well as AFGL~2136 and AFGL~2591 \citep{barr22}. While iSHELL measurements \citep{li22, barr20} provide a better {constraint} on the amount of gaseous CO from the same region, we derive high relative abundances of H$_2$O to CO; {i.e.}, $\sim1$ to 1.5 for \wt, 1.6 for AFGL~2136, and 7.4 for AFGL~2591. Such a high relative H$_2$O to CO abundance is expected for warm, dense gas where gas phase chemistry rapidly converts the available O not in CO into H$_2$O \citep{kn96}. As a comparison, these values are much higher than the H$_2$O/CO = $10^{-4}$ derived from submillimeter observations by \textit{Herschel}-HIFI toward the hot core region of AFGL~2591~\citep{kaz14}, or the value of 4.4\% from the warm 200~K component identified in this water study (see Table~\ref{table:sum-prop}). On the other hand, a high relative H$_2$O to CO abundance from 1 to 2 was observed toward T Tauri and Herbig disks \citep{cn08, salyk11}. 

In the cold dense ISM, there is a well-documented problem of the missing oxygen budget \citep{whittet10}. While \citet{h09} predicted that oxygen not in silicates or oxides should be eventually converted into gaseous CO and ices, a substantial shortfall of oxygen is observed. In the study of the Taurus complex dark clouds \citep{whittet10}, the combined contributions of gaseous CO, ice, and silicate/oxide account for less than 300~ppm of the elemental oxygen compared to the solar value of 490~ppm \citep{asplund09}. We observed a similar missing oxygen reservoir in \wt: assuming 284~ppm of the O in diffuse clouds~\citep{cartledge04}, we only see a value of 58.1~ppm (20.4\%) in cold, dense clouds~(see Table~\ref{table:abun}). {If one uses local B stars as the interstellar standard, the total budget of the elemental abundance of oxygen is even higher~\citep[575 ppm rather than 490 ppm,][]{np12}. }Therefore, a budget close to the oxygen abundance in silicate \citep[$\sim200$~ppm, ][]{tielens87} is missing or is locked up in an unidentified form, which is referred to as ``the unidentified depleted oxygen (UDO)" in \citet{whittet10}. 

The intrinsic properties of the reservoir of the missing oxygen remain mysterious. Refractory dust compounds like carbonates are implausible, as they will also survive and appear in the diffuse medium. Neither gas-phase nor solid-phase O$_2$ are possible as well. While SWAS observations~\citep{goldsmith00} towards massive protostars provide an upper limit of 0.1~ppm on the gaseous O$_2$, solid O$_2$ is too volatile in the line of sight of \wt. Oxygen-bearing organics in solid ice would be a potential carrier, although no significant detection of the related bonds has yet been detected at infrared wavelengths~\citep{gibb04}. We suggest that the MIRI spectrograph on board the \textit{JWST} is well-suited to study such an organic inventory. {One other possibility of the UDO is a population of very large water ice grains ($>1~\mu$m) in the cold gas~\citep{jenkins09}. These large grains produce gray extinction and their putative presence is hard to refute. }

Similar to the ``oxygen crisis", a depletion problem in elemental carbon exists in the envelope of \wt~(see Table~\ref{table:abun}). The total amount of carbon (32~ppm) comprises only 19.7\% of the value expected in diffuse clouds~\citep[160~ppm, ][]{cardelli96, sofia97}. As discussed in \citet{li22}, one speculation is that the carbon-containing ice compounds were converted into an organic residue by prolonged UV photolysis \citep{bs95, bs97, vino13}.

{Both the ``oxygen crisis" and the ``carbon crisis" were observed in AFGL~2136 and AFGL~2591 as well. In contrast to previous studies that relied on comparing pencil beam IR absorption line studies with sub-millimeter emission observations, the IR pencil beam samples the same material in absorption for these massive protostars.} As shown in Table~\ref{table:abun}, the depletion problems are more severe for AFGL~2591, but less severe for AFGL~2136. We suggest that further studies of the different oxygen reservoirs could help pinpoint the processes involved in the missing oxygen or carbon reservoirs by studying a large enough sample with diverse characteristics.

\section{Summary}\label{sec:sum}

We conducted high spectral resolution ($R$=50,000; 6~\kms) spectroscopy from 5--8 $\mu$m with EXES on board SOFIA toward the hot core region associated with the massive binary protostar \wt. By comparing with the LTE models constructed with the existing laboratory line information, we identified about 180 $\nu_2=1-0$ and 90 $\nu_2=2-1$ absorption lines. Preliminary Gaussian fittings and rotation diagram analyses reveal two hot components with $T>$ 600~K and one warm component of 190~K. However, the large scatter in the rotation diagrams of the two hot components reveals 1) opacity effects, 2) that the absorption lines are not optically thin, and 3) the total column densities derived from the rotation diagrams are underestimated. 

We adopted two curve-of-growth analyses to account for the opacity effects of the hot components. One model considers absorption in a foreground slab partially covering the background emission. The other model assumes absorption in the photosphere of a circumstellar disk with an outward-decreasing temperature in the vertical direction. In both models, about half of the data points converted from the $\nu_2=1-0$ transitions are located on the logarithmic part of the curve-of-growth, confirming that the corresponding absorption lines are optically thick. The two curve-of-growth analyses correct the column densities by at least an order of magnitude and lower the derived excitation temperatures accordingly. We note that for the disk model, the results of the curve-of-growth analysis depend on the adopted velocity width $\sigma_v$ and the parameter $\epsilon$ that characterizes absorption and scattering of the absorption line. We provide a reference table for different $\sigma_v$ and $\epsilon$ as the two parameters are poorly constrained.

Although our SOFIA-EXES observations do not spatially resolve the binary protostars in W3 IRS 5, using the kinematic and temperature characteristics, we link each H$_2$O component to a spatially separate CO component identified in IRTF/iSHELL observations~($R$=88,100). Specifically, the warm H$_2$O component `W' is linked to the shared warm CO component MIR1-W1/MIR1-W2, the hot H$_2$O component `H1' is linked to the MIR2-H2 in CO, and the hot H$_2$O component `H2' is considered to be related to  one of the CO MIR1-W$^\prime$ components. 

{Once the connections of H$_2$O components and CO components were established, we discussed the physical origins of H$_2$O components in light of a better understanding of CO components.} From our analysis, we conclude that the disk model is the preferred one over the slab model out of the considerations of the geometry constraints, although disk models face challenges in realizing such an internal heating mechanism. 

We derive the H$_2$O/CO abundance ratio based on the results of the disk model and discuss the chemical abundances along the line of sight based on the H$_2$O-to-CO connection. For the hot gas, we derive a high H$_2$O/CO abundance ratio of 0.9. Such a high relative H$_2$O to CO abundance is expected for warm, dense gas where gas phase chemistry rapidly converts the available O not in CO into H$_2$O. For the cold gas, we observe a substantial shortfall of oxygen in agreement with earlier studies of cold dense clouds. We suggest that organics in solid ice are the potential carrier.

We greatly thank the anonymous referee for his/her thorough thoughts and helpful suggestions to improve the quality of this paper. We acknowledge the support for the EXES Survey of the Molecular {Inventory} of Hot Cores (SOFIA $\#$08-0136) at the University of Maryland from NASA (NNA17BF53C) Cycle Eight GO Proposal for the Stratospheric Observatory for Infrared Astronomy (SOFIA) project issued by USRA. 

\software{Astropy \citep{astropy13, astropy18}, Matplotlib \citep{hunter07}, NumPy \citep{harris20}, SciPy \citep{virtanen20}.}

\appendix

\section{Telluric Molecular Lines from 5.36 to 7.92~$\mu$m}\label{app:tl}

\begin{figure*}[!h]
    \centering
    \includegraphics[width=\linewidth]{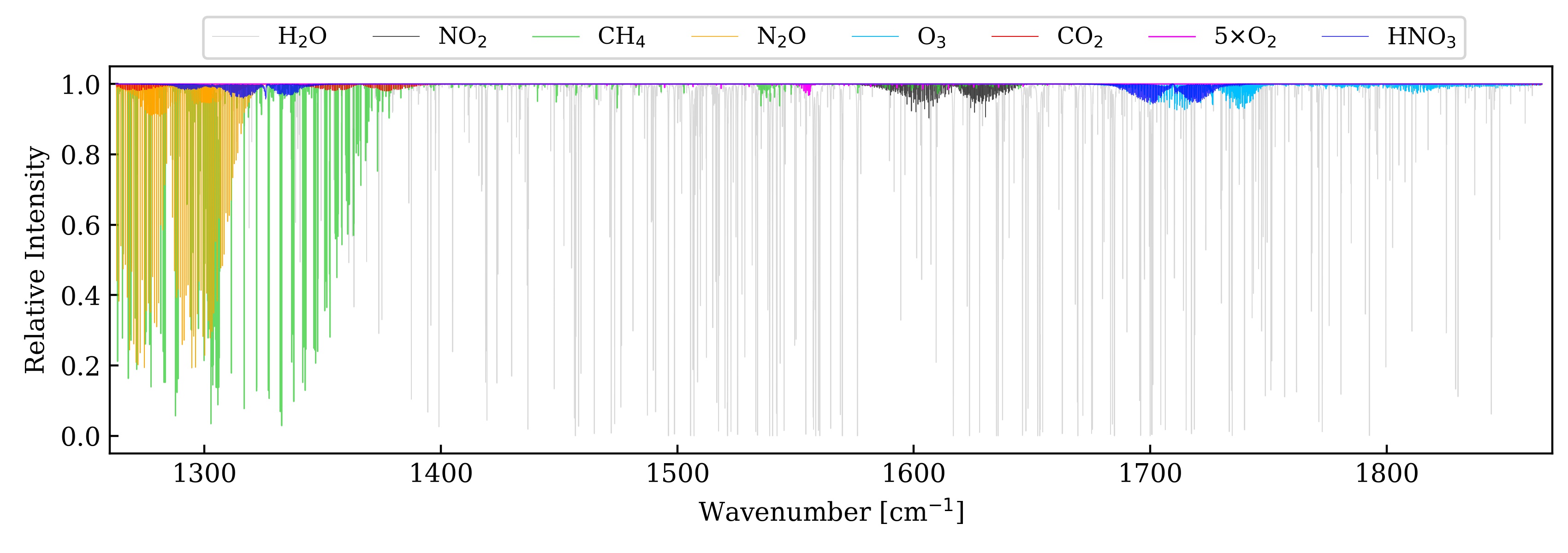}
    \caption{Important atmospheric telluric lines from 5.36 to 7.92~\um.}
    \label{fig:telluric}
\end{figure*}

We present in Figure~\ref{fig:telluric} the important telluric lines from 5.36 to 7.92~\um\ produced by the PSG models under representative observational parameters (see Table~\ref{table:psg}). Eight molecular species, \hto, \oth, \chf, \notwo, \ntwoo, \cotwo, HNO$_3$, and \otwo\ are among the most important ones. For the 15 observational settings, the input surface pressure and scaling factors are tuned {by hand} to visually match the telluric features in observed spectra, and therefore, may not represent the actual values.

\begin{deluxetable*}{ccc rrrr rrrr}[!t]
\tablecolumns{11}
\tabletypesize{\scriptsize}
\tablecaption{Inputs for the PSG Models \label{table:psg}}
\tablehead{\colhead{Source} & \colhead{$\lambda$} & \colhead{$P$} & \colhead{\hto}  & \colhead{\oth} & \colhead{\chf} & \colhead{\notwo} & \colhead{\ntwoo} & \colhead{\cotwo} & \colhead{HNO$_3$} & \colhead{\otwo}       \\ & (\um) &(bar) &&&&&\\
 \colhead{(1)} & \colhead{(2)}  & \colhead{(3)} & \colhead{(4)} & \colhead{(5)} & \colhead{(6)} & \colhead{(7)} & \colhead{(8)}& \colhead{(9)} & \colhead{(10)} & \colhead{(11)}}
\startdata
\wt&	5.36--5.51	&	0.75	&	1	&	1	&	--	&	--	&	--	&	--	&	--	&	--	\\
&	5.48--5.67	&	0.7	&	1	&	1	&	--	&	--	&	--	&	--	&	--	&	--	\\
&	5.65--5.84	&	0.85	&	0.7	&	1	&	--	&	--	&	--	&	--	&	--	&	--	\\
&	5.83--6.02	&	0.92	&	0.5	&	1.5	&	--	&	--	&	--	&	--	&	2	&	--	\\
&	6.01--6.20	&	0.83	&	1	&	--	&	1	&	1	&	--	&	--	&	--	&	--	\\
&	6.01--6.20	&	0.75	&	1	&	--	&	1	&	1.5	&	--	&	--	&	--	&	--	\\
&	6.18--6.37	&	0.83	&	0.8	&	--	&	1	&	1.3	&	--	&	--	&	--	&	--	\\
&	6.19--6.37	&	0.85	&	0.6	&	--	&	1	&	0.8	&	--	&	--	&	--	&	--	\\
&	6.35--6.61	&	0.75	&	0.8	&	--	&	2	&	--	&	--	&	--	&	--	&	1	\\
&	6.35--6.61	&	0.83	&	0.9	&	--	&	1.5	&	--	&	--	&	--	&	--	&	--	\\
&	6.59--6.85	&	0.77	&	1	&	--	&	2	&	--	&	--	&	--	&	--	&	--	\\
&	6.59--6.85	&	0.85	&	1	&	--	&	1	&	--	&	--	&	--	&	--	&	--	\\
&	6.79--7.06	&	0.75	&	1.5	&	--	&	1.8	&	--	&	--	&	--	&	--	&	--	\\
&	7.19--7.45	&	0.8	&	1	&	--	&	1.5	&	--	&	--	&	1	&	--	&	--	\\
&	7.67--7.92	&	0.7	&	1	&	--	&	2	&	--	&	2.5	&	1	&	1	&	--	\\
Sirius &	7.18--7.46	&	0.7	&	1.4	&	--	&	1.7	&	--	&	--	&	1	&	--	&	--	\\
\enddata
 \tablecomments{Column (1): Sirius is the standard star for the observation session on 2022-02-24 at 7.19--7.45~\um. Column (3): the input surface pressure (Earth at 4084~m) in the PSG models. Column (4)--(11): the input scaling factors of different atmospheric molecular species in the PSG models. See Figure~\ref{fig:telluric} for an illustration of the contribution of different molecular species at different wavelengths.}
\end{deluxetable*} 

\clearpage

\section{Data Reduction: Sirius vs. PSG models}\label{app:sirius}

\begin{figure}[!t]
    \centering
    \includegraphics[width=\linewidth]{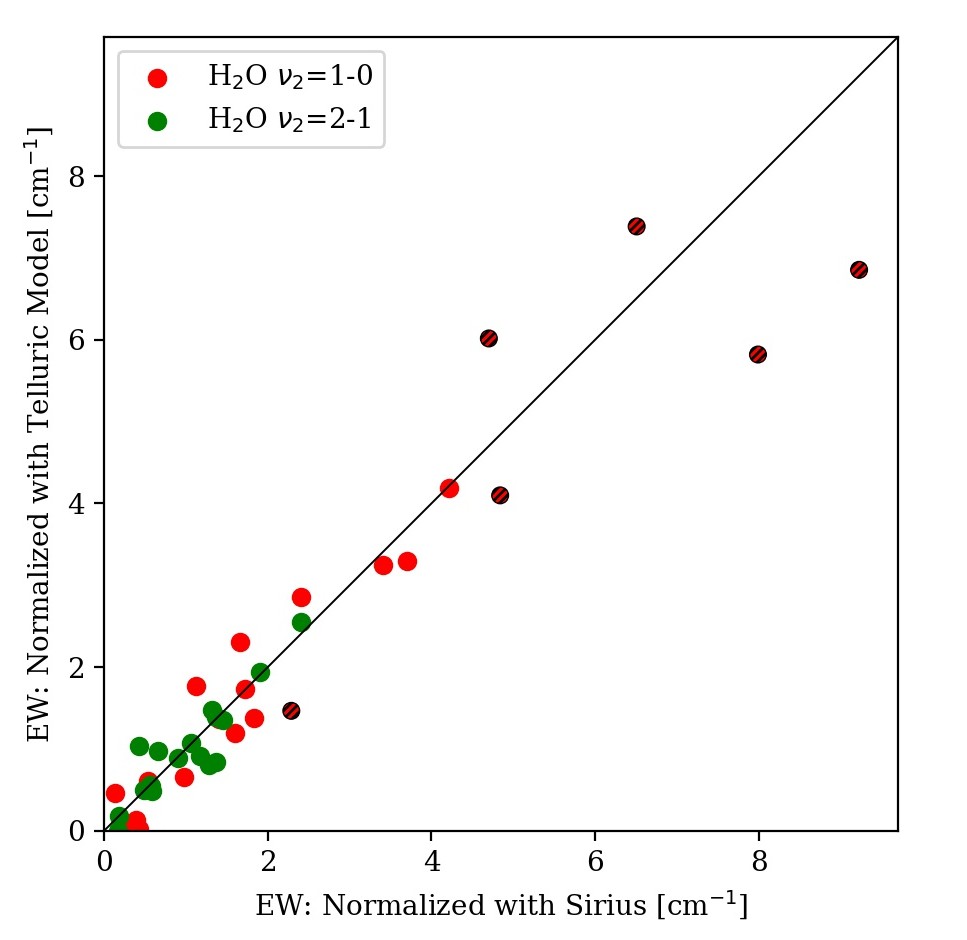}
    \caption{Comparison between the equivalent widths determined from spectra calibrated with the standard star Sirius versus using PSG models for the telluric correction.}
    \label{fig:sirius}
\end{figure}

The standard star, Sirius, was observed only in one setting from 7.28--7.46~\um. While the Sirius spectrum has advantages in reflecting the actual baseline, we compare the qualities of the results derived by using the Sirius spectrum as well as using the PSG models. The data reduction with Sirius is straightforward, as the median filtering processes or the modeling of telluric lines are not needed.

As a result, we present in Figure~\ref{fig:sirius} the equivalent widths of each individually identified line derived from the two data reduction methods. We conclude that the results are in good agreement with each other by $\sim10\%$.

\section{The Hidden Component at $-38$~\kms}\label{section:hidden}

As presented in \S~\ref{sec:da} and \S~\ref{sec:re}, data analyses in this paper are based on the decomposition of the absorption profiles into the three components at $-54.5, -45$, and $-39.5$~\kms\ in low-energy levels, while the $-39.5$~\kms\ component is hot. However, we realize that one cold component of $\sim$50~K may exist at $-38$~\kms\ because of the detection of the cold CO component~\citep{li22}. 

\begin{figure}
    \centering
    \includegraphics[width=\linewidth]{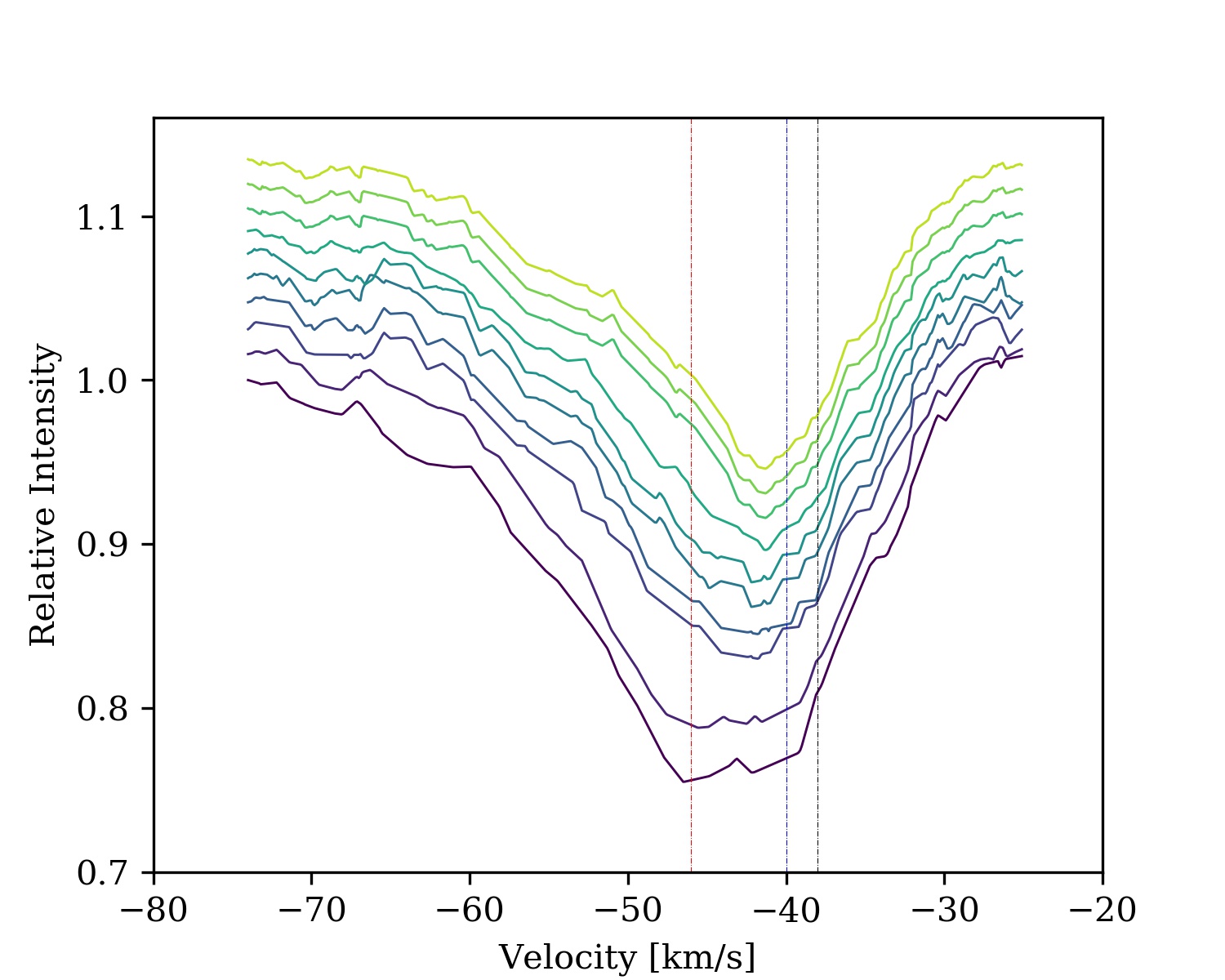}
    \caption{Accumulative average spectra from 100 to 1000 K. Each spectrum represents the median of spectra in energy levels between 0 to 100 K, 0 to 200~K, ..., 0 to 1000~K. The dashed vertical lines represent $-46$, $-40$, and $-38$~\kms.}
    \label{fig:avg_1000}
\end{figure}

\begin{figure}
    \centering
    \includegraphics[width=0.95\linewidth]{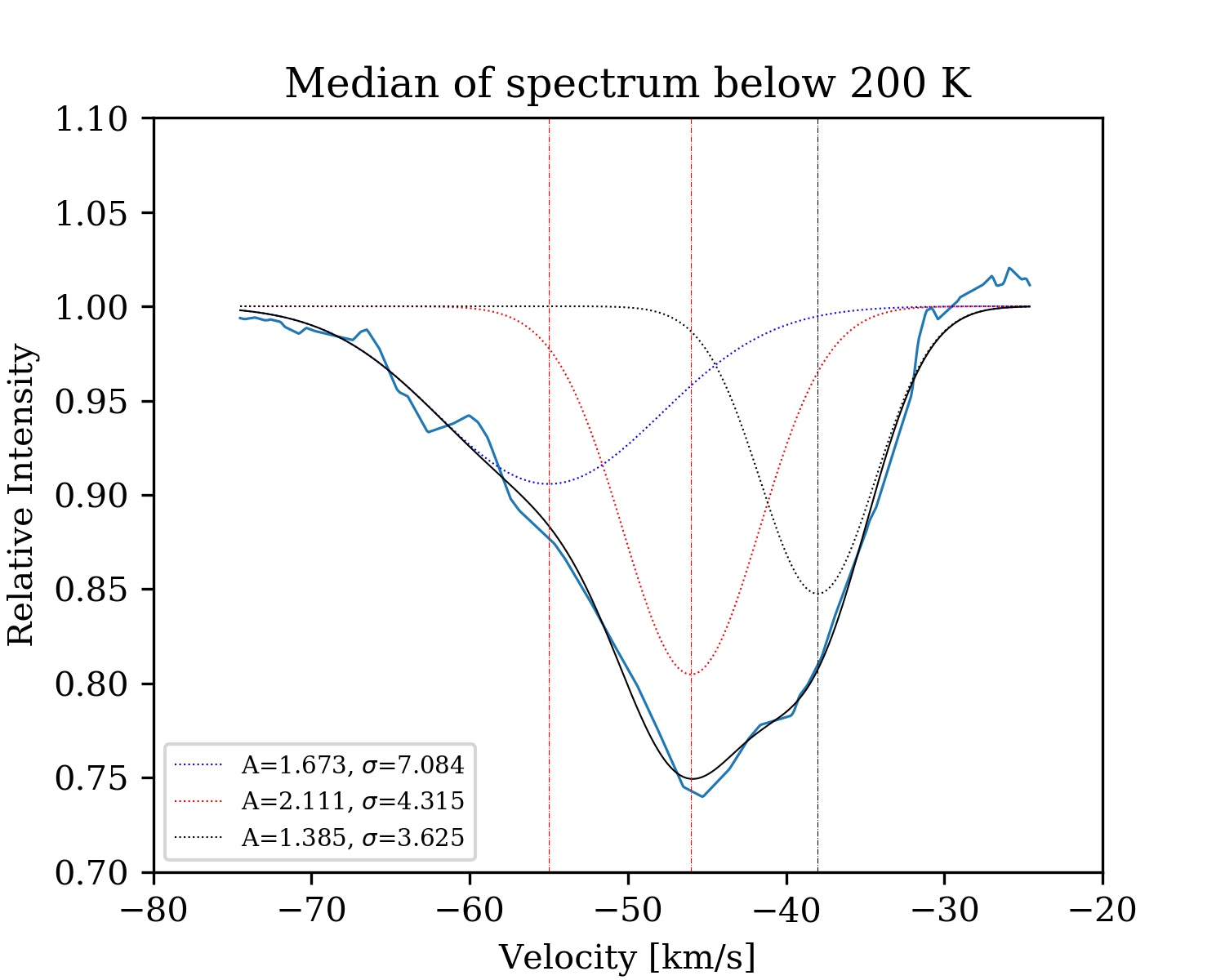}
    \includegraphics[width=0.95\linewidth]{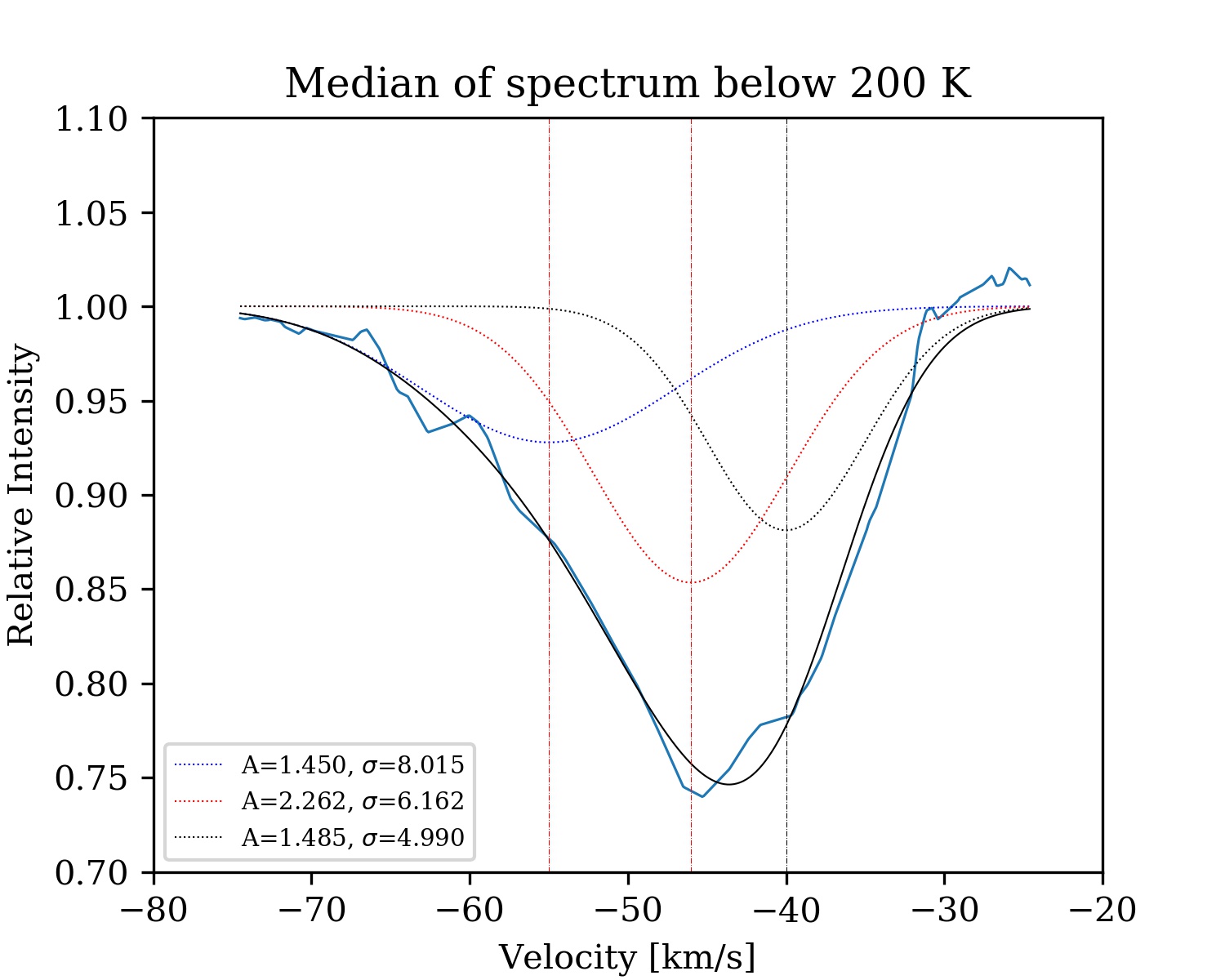}
    \caption{Two different Gaussian fitting results for the spectrum averaged below 200~K. In the upper panel, the central velocities are $-55, -46$, and $-38$~\kms. In the lower panel, the central velocities are $-55, -46$, and $-40$~\kms.}
    \label{fig:avg_200}
\end{figure}

We argue that a cold component is possibly hidden in Figure~\ref{fig:avg_1000}, in which a list of accumulative average spectra is presented. We note that the absorption feature at $-35$~\kms\ is possibly related to the cold component because it disappears as the energy levels increase. We also present two different Gaussian fitting methods for the average spectrum below 200~K in Figure~\ref{fig:avg_200}, while in one, we fix the right wing with a component at $-38$~\kms\ and in the other at $-40$~\kms. We conclude that the previous one provides a better fit, suggesting that at this energy level, the cold $-38$~\kms\ component possibly dominates the line profile. 

\begin{figure}
    \centering
    \includegraphics[width=0.95\linewidth]{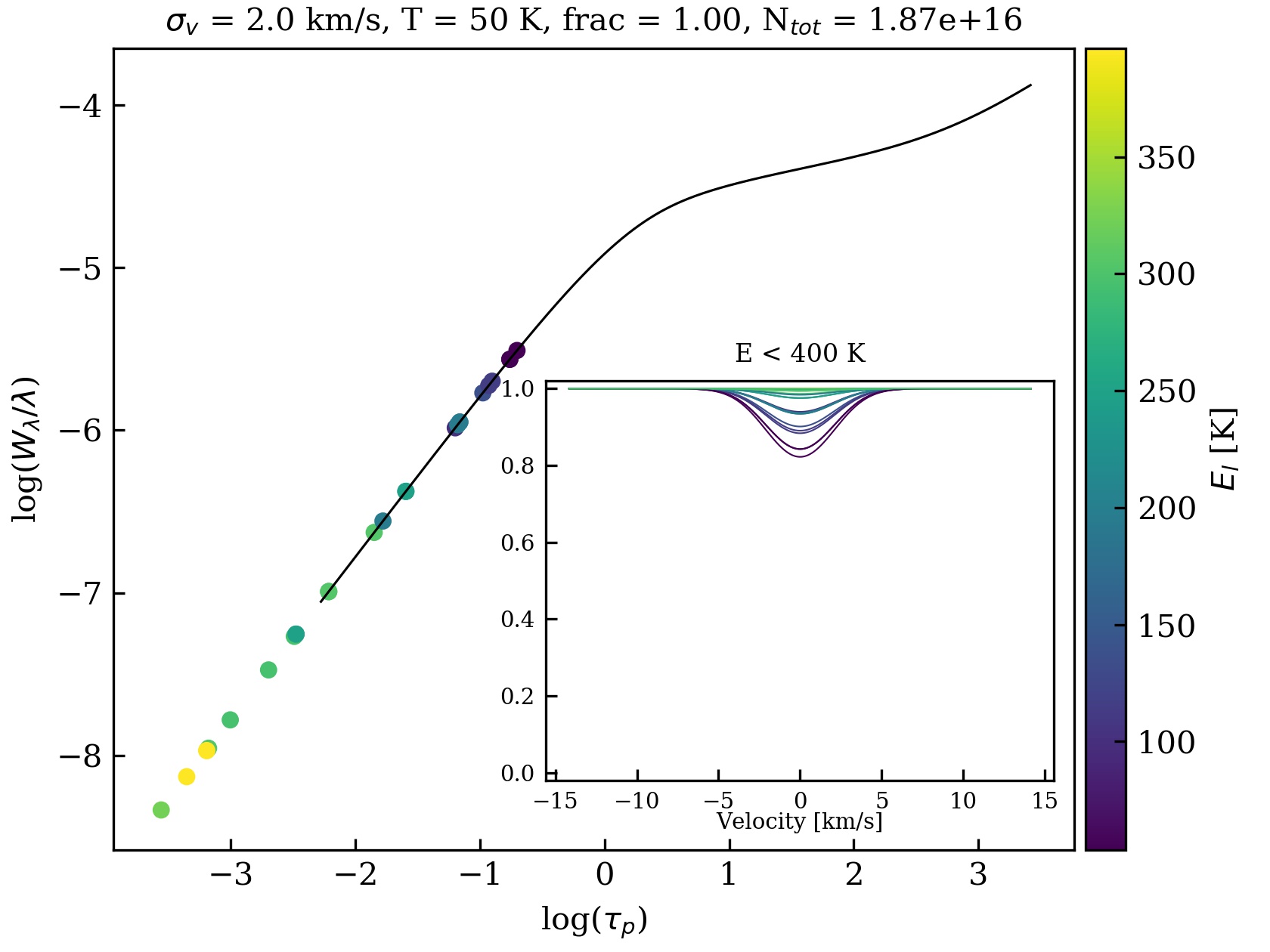}
    \caption{Expected cold line profiles of water and the correspondent curve-of-growth based on parameters constrained in CO observations.}
    \label{fig:11}
\end{figure}

Even if the cold $-38$~\kms\ component does exist, we conclude that the water-to-CO abundance is still low at this temperature. Otherwise, very saturated water absorption lines will dominate the line profiles. We estimate an upper limit of such a water-to-CO ratio of 0.4$\%$ by Figure~\ref{fig:11}.

\section{Additional Figures}\label{app:e}

\begin{figure*}
    \centering
    \includegraphics[width=0.47\linewidth]{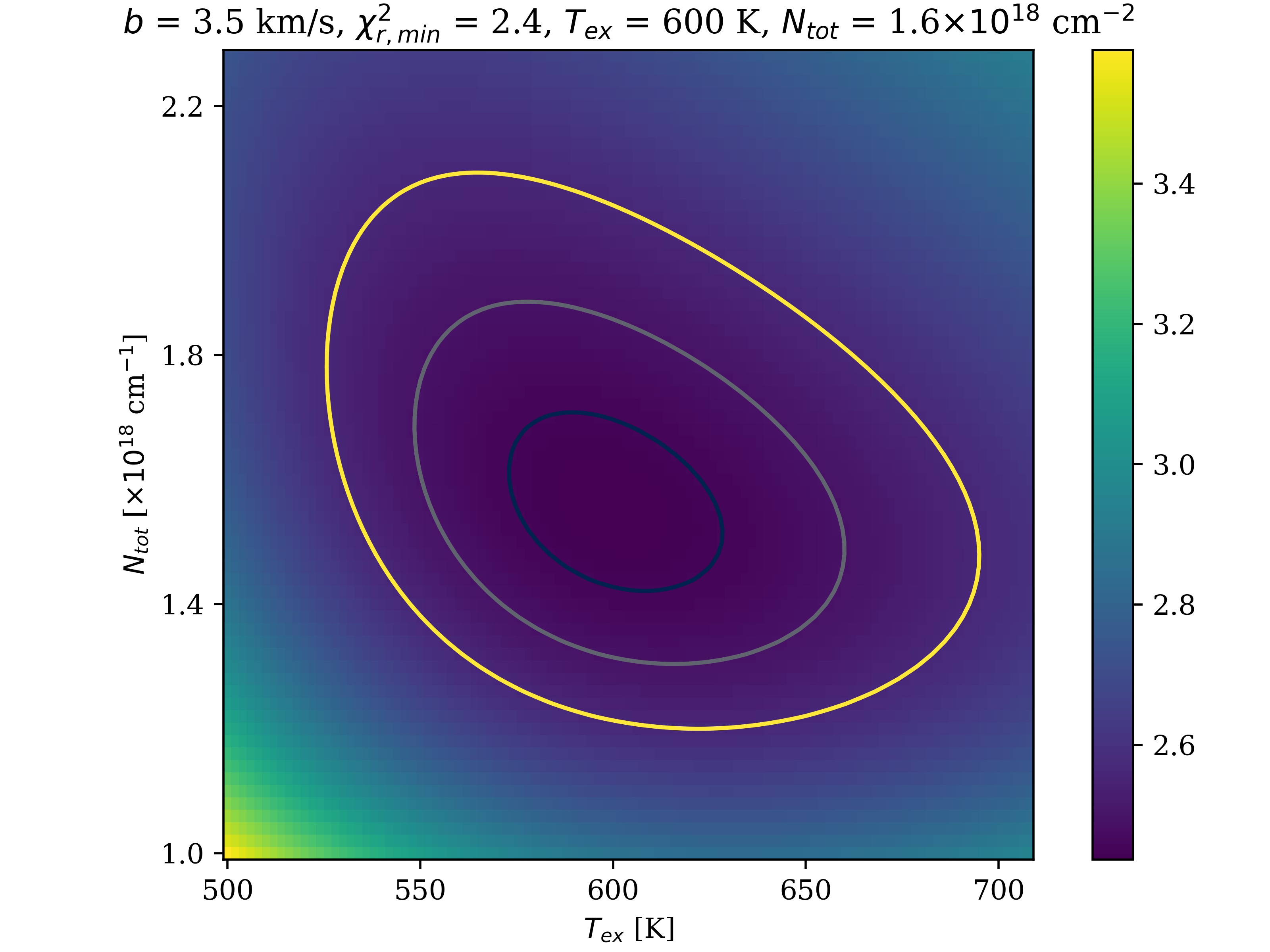}
    \includegraphics[width=0.45\linewidth]{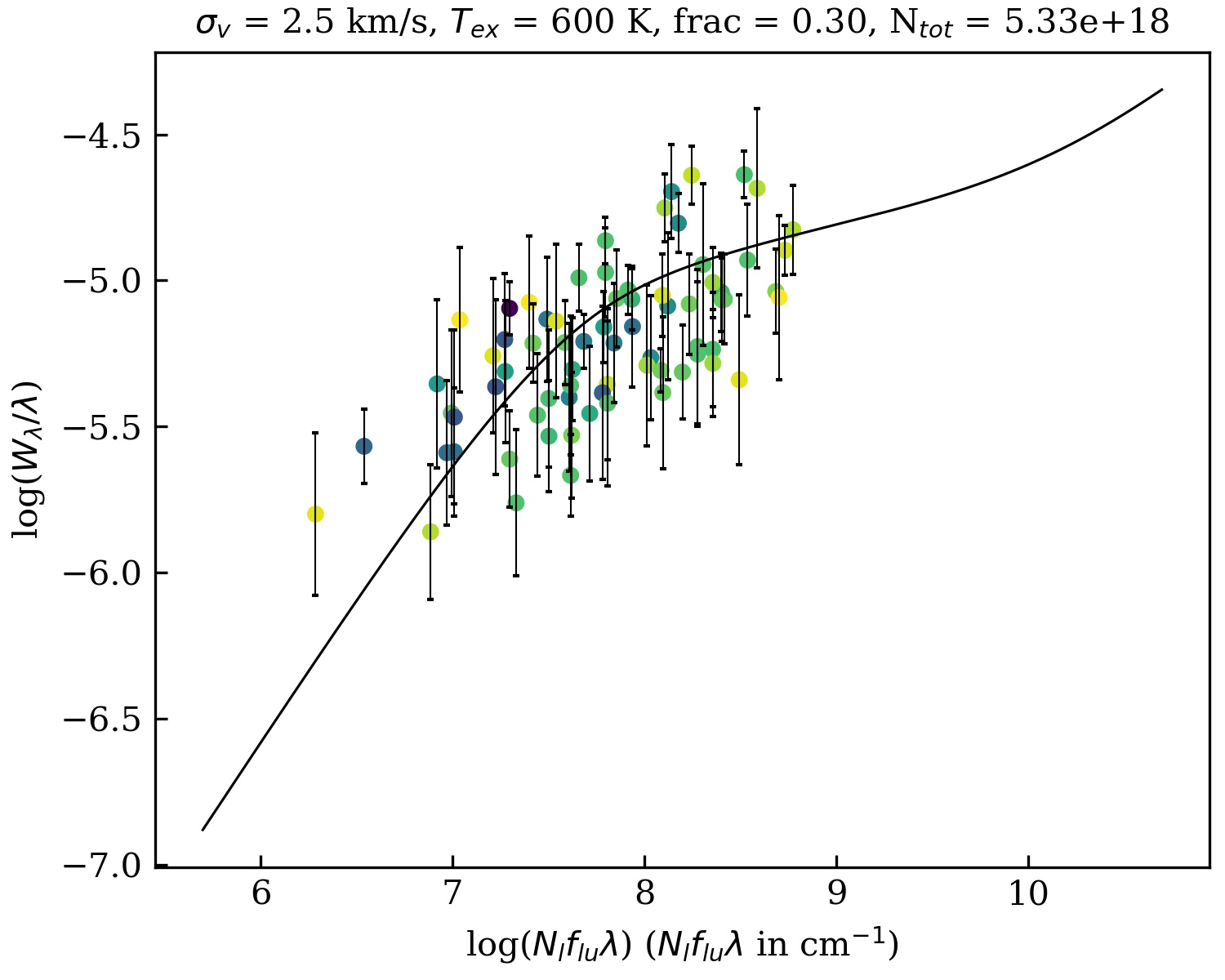}\\
    \includegraphics[width=0.47\linewidth]{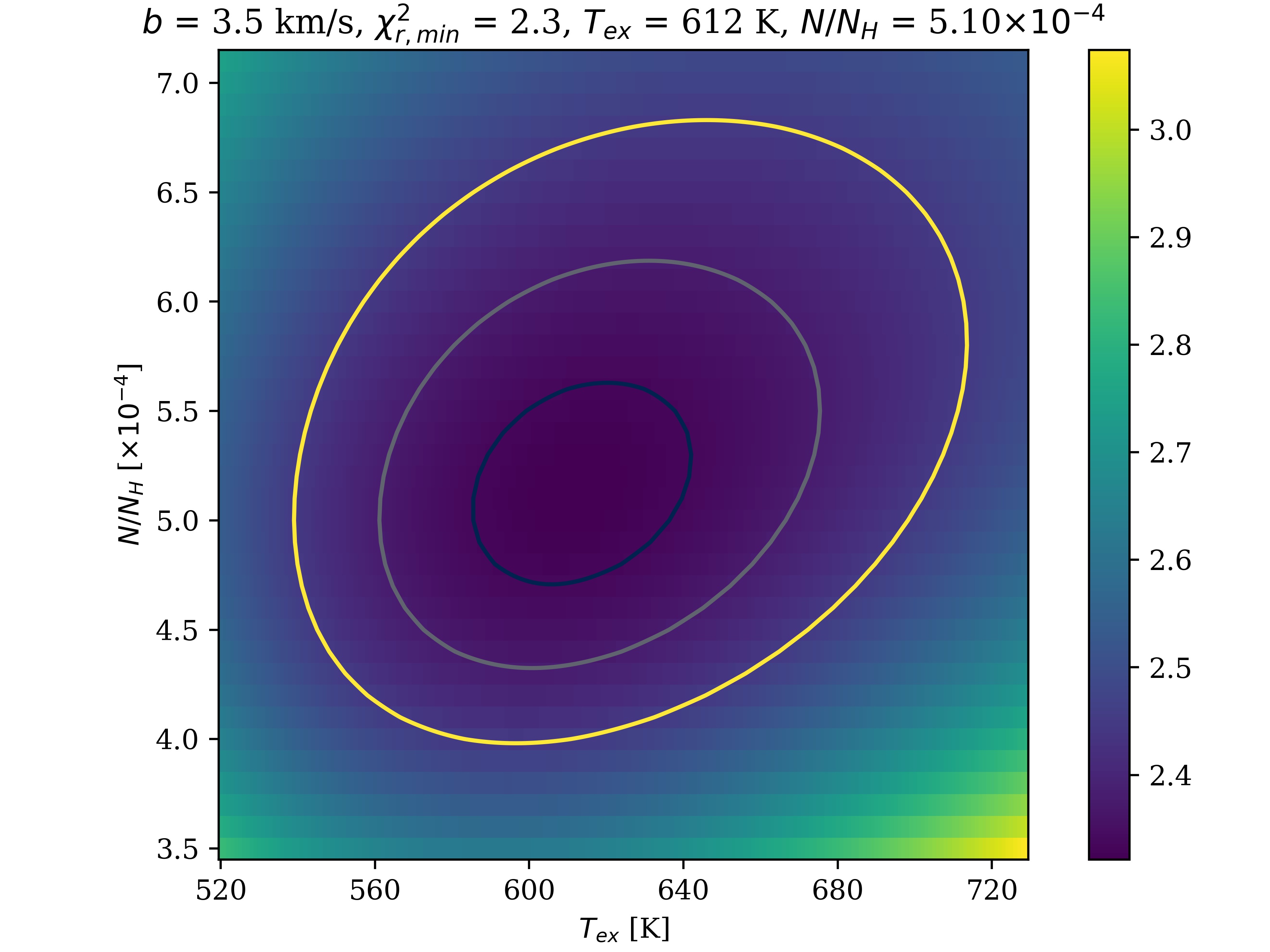}
    \includegraphics[width=0.45\linewidth]{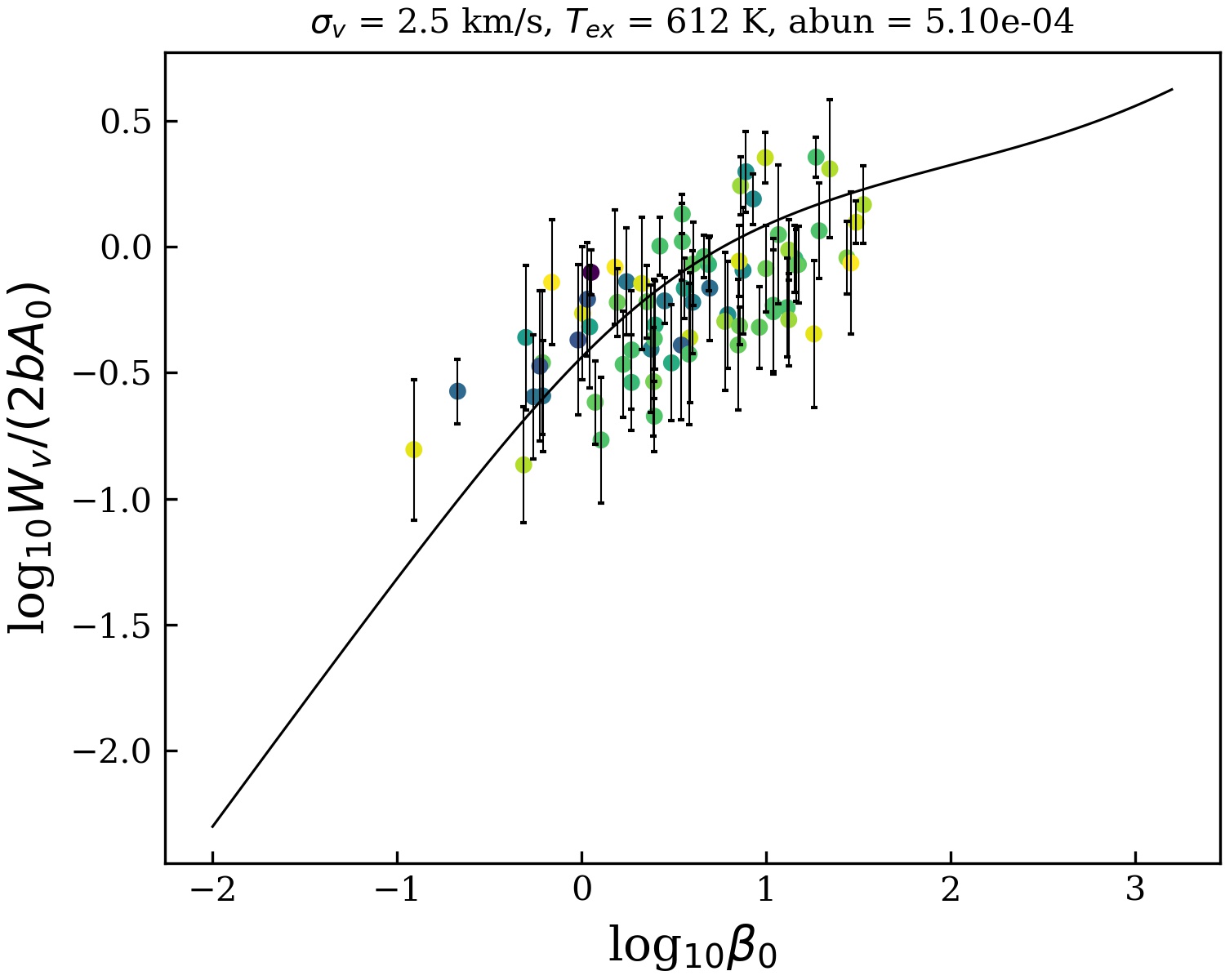}
    \caption{\textit{Left panels:} Grid-search results of `H2' ($\nu_2$=2--1 transition) for both the slab and the disk model illustrating the best-fitting results. The contours represent the 1$\sigma$, 2$\sigma$, and 3$\sigma$ uncertainty levels. \textit{Right panels}: the curves of growth for the slab and the disk model.}
    \label{fig:cog-h2}
\end{figure*}

Figure~\ref{fig:cog-h2} presents the two grid-search results and the best-fitted curve-of-growth for the `H2' component.

\begin{figure*}
    \centering
    \includegraphics[width=0.47\linewidth]{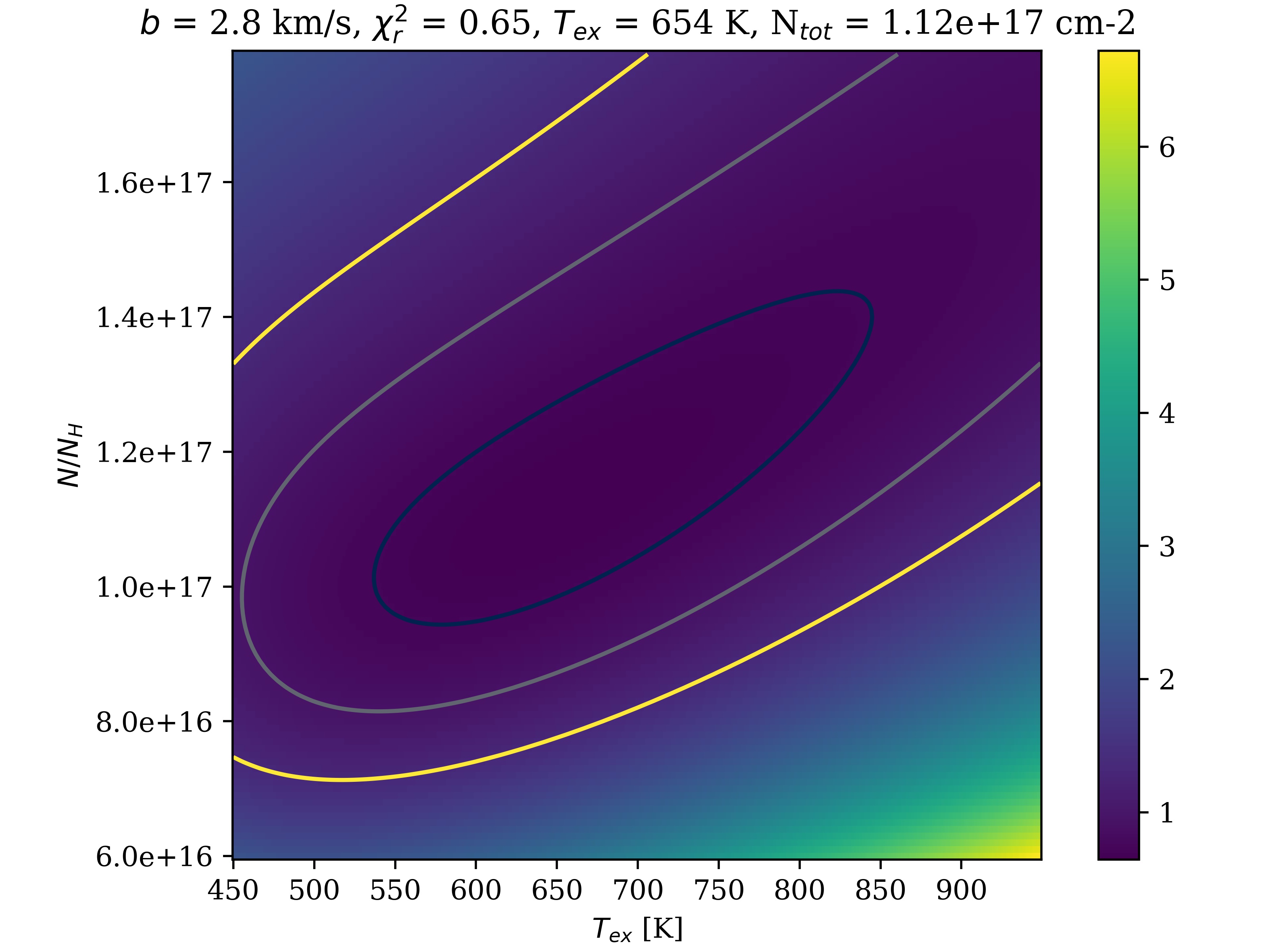}
    \includegraphics[width=0.45\linewidth]{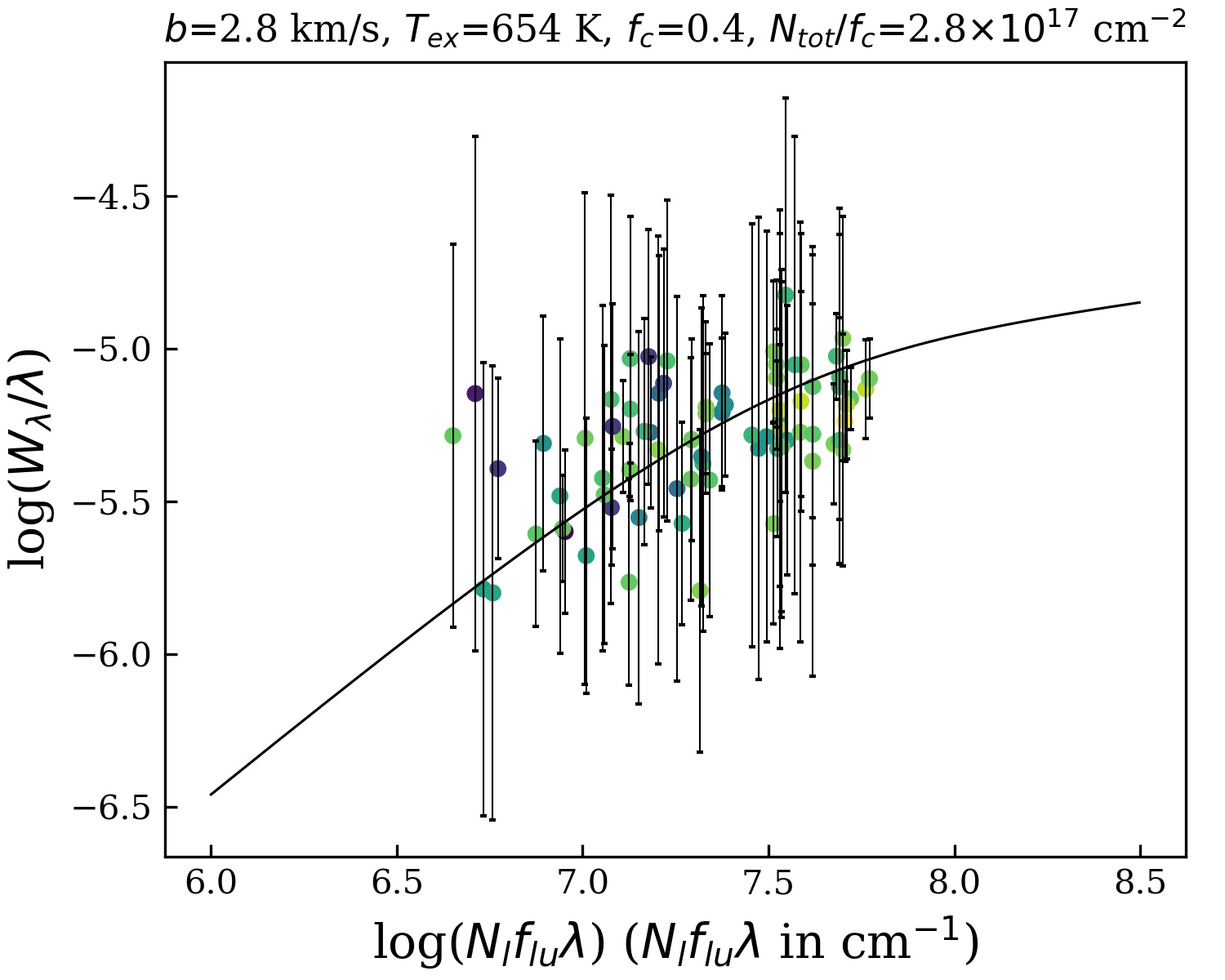}\\
    \includegraphics[width=0.47\linewidth]{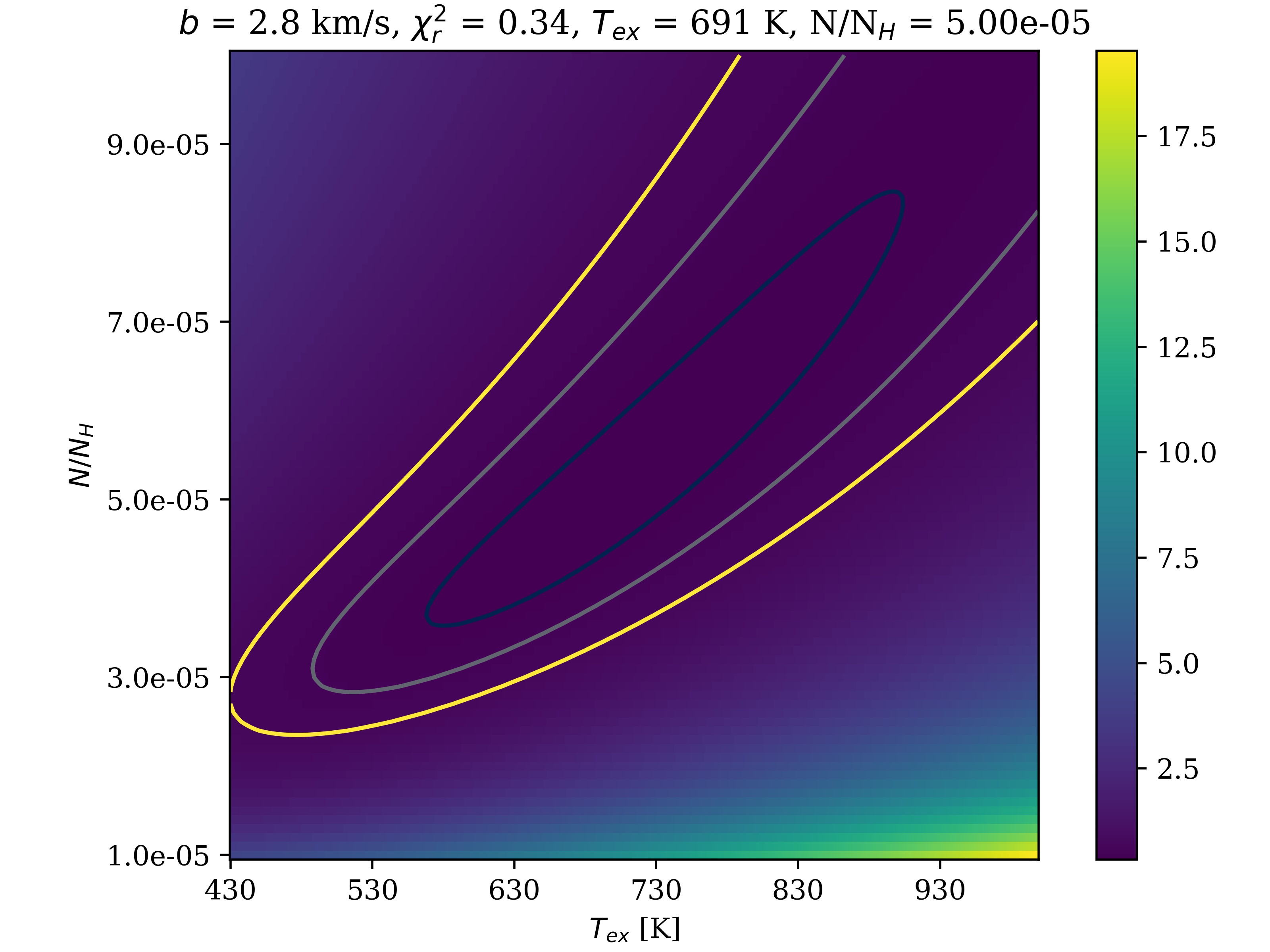}
    \includegraphics[width=0.45\linewidth]{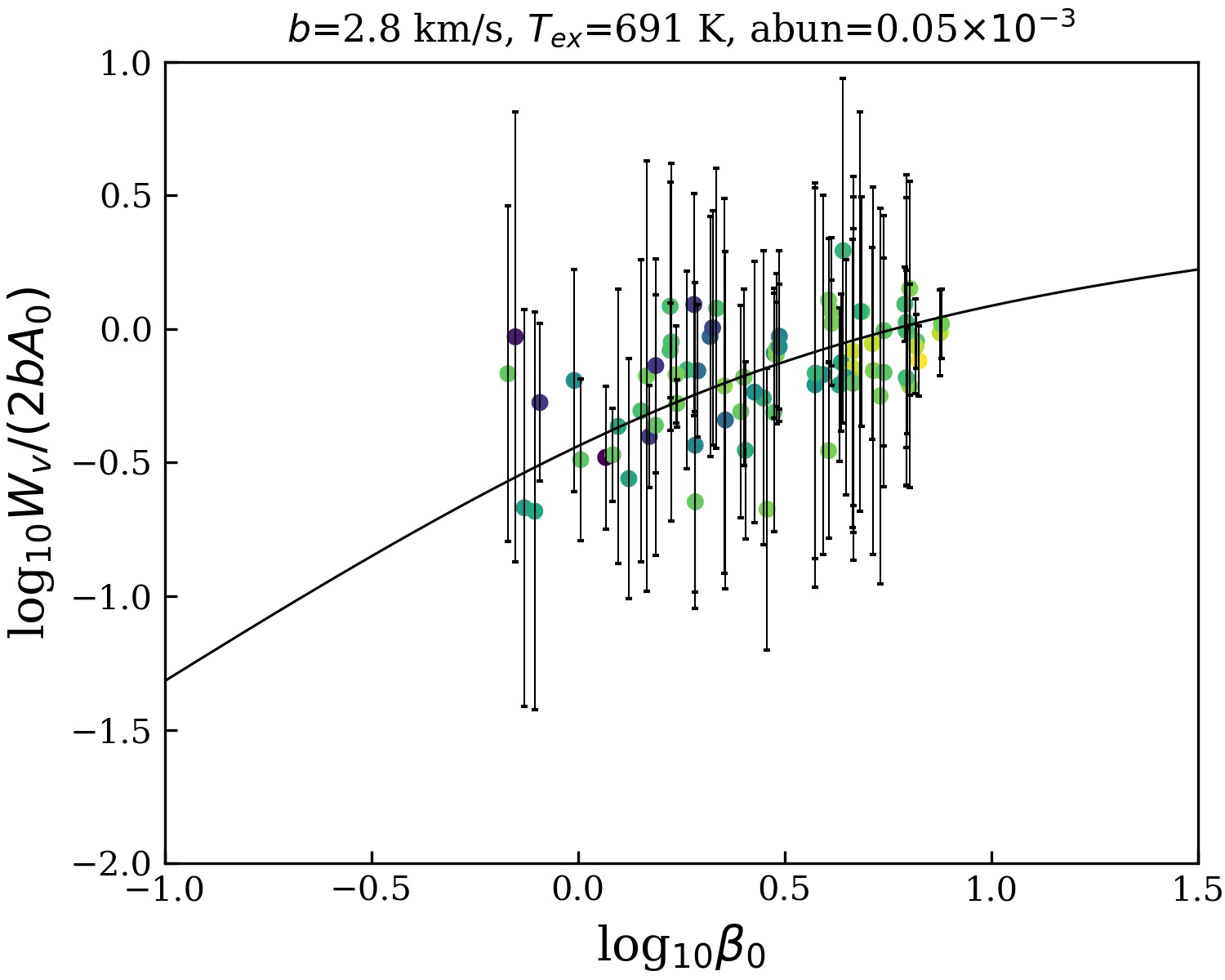}
    \caption{\textit{Left panels:} Grid-search results of `H1' ($\nu_2$=2--1 transition) for both the slab and the disk model illustrating the best-fitting results. The contours represent the 1$\sigma$, 2$\sigma$, and 3$\sigma$ uncertainty levels. \textit{Right panels}: the curves of growth for the slab and the disk model.}
    \label{fig:cog-h1-2-1}
\end{figure*}

\section{List of the Water Lines}\label{app:table}

We present from Table~\ref{table:8} to \ref{table:12} the properties of decomposed water lines, \hto\ \nuu$_2$=1--0 and \nuu$_2$=2--1, of \wt\ from 5.36 to 7.92~\um\ in this study. For `H2' in the $\nu_2$=1--0 transitions, the velocity centers of lines with $E_l <$ 800 K are fixed at $-54$~\kms. {For `W', in the $\nu_2$=1--0 transitions, the velocity centers are fixed at $-45$~\kms. $\tau_{{p}, \textrm{thin}}$, $\tau_{{p}, \textrm{slab}}$, and $\beta_0$ are calculated from the best-fitted temperatures and column densities from the rotation diagram, the slab model, and the disk model. }

The Gaussian fitting results of the central velocity, \vLSR, vary among different transitions of the same physical component. The dispersion is contributed by both the uncertainty in \vLSR\ as well as the uncertainty level of the fitting results, $\Delta$\vLSR. We report the fitted central velocity and the uncertainty level of one component as follows:

\begin{equation}
    v_{LSR} = \frac{1}{N}\sum_i v_{LSR, i},
\end{equation}
and
\begin{equation}
\Delta v_{LSR} = \sqrt{\frac{1}{N}\sum_i (\Delta v_{LSR, i})^2 + \left\langle v_{LSR, i}^2 \right\rangle - \left\langle v_{LSR, i} \right\rangle^2}.
\end{equation}

\startlongtable
\begin{deluxetable*}{l c r c c r c c c cccc}\label{table:8}
\tablecolumns{13}
\tablecaption{Line Parameters of the $-39.5$~\kms\ Component (`H1') in the $\nu_2$=1--0 Transition}
\tablehead{\colhead{Transition}  & \colhead{$\lambda$} 
& \colhead{$E_l$} & \colhead{$g_l$} & \colhead{$g_u$} & \colhead{$A_{ul}$} & \colhead{\vLSR} & \colhead{$\sigma_v$}  & \colhead{$N_l$} & \colhead{$W_v$} & \colhead{log$_{10}(\tau_{p, \textrm{thin}})$} & \colhead{log$_{10}(\tau_{p, \textrm{slab}})$} & \colhead{log$_{10}(\beta_0)$}  \\ & \colhead{(\um)} & \colhead{(K)} &  & & \colhead{(s$^{-1}$)} & \colhead{(\kms)} & \colhead{(\kms)} & \colhead{($\times 10^{15}$~cm$^{-2}$)} & \colhead{(\kms)}  & & & }
\startdata
12$_{1,11}$-11$_{2,10}$	&	5.3737	&	2193.9	&	23	&	25	&	10.9	&	-39.2	$\pm$	1.3	&	6	$\pm$	1	&	5.6	$\pm$	0.5	&	3.67	$\pm$	0.802	&	-0.55	&	0.35	&	0.58	\\
11$_{3,8}$-10$_{4,7}$	&	5.3772	&	2274.8	&	63	&	69	&	3.3	&	-40.7	$\pm$	1.4	&	5.2	$\pm$	1	&	11.4	$\pm$	1.6	&	2.372	$\pm$	0.7	&	-0.68	&	0.2	&	0.43	\\
7$_{6,1}$-7$_{5,2}$	&	5.4183	&	1524.6	&	45	&	45	&	1.3	&	-40.5	$\pm$	0.8	&	4.2	$\pm$	0.8	&	25.5	$\pm$	3.3	&	1.904	$\pm$	0.526	&	-0.7	&	0.3	&	0.5	\\
11$_{2,10}$-10$_{1,9}$	&	5.4248	&	1860	&	21	&	23	&	10.2	&	-41.4	$\pm$	0.6	&	6	$\pm$	0.5	&	8.1	$\pm$	0.4	&	4.98	$\pm$	0.53	&	-0.39	&	0.61	&	0.82	\\
11$_{6,5}$-11$_{5,6}$	&	5.4521	&	2875.6	&	69	&	69	&	2.4	&	-39	$\pm$	1.2	&	4	$\pm$	1.2	&	4	$\pm$	0.6	&	0.6	$\pm$	0.18	&	-1.21	&	-0.48	&	-0.23	\\
10$_{1,9}$-9$_{2,8}$	&	5.4862	&	1554.1	&	19	&	21	&	9.5	&	-41	$\pm$	0.8	&	6	$\pm$	0.8	&	7.3	$\pm$	0.5	&	4.51	$\pm$	0.724	&	-0.44	&	0.84	&	1.04	\\
10$_{1,9}$-9$_{2,8}$	&	5.4862	&	1554.1	&	19	&	21	&	9.5	&	-38.2	$\pm$	0.8	&	5.6	$\pm$	0.9	&	6.1	$\pm$	0.4	&	3.766	$\pm$	0.592	&	-0.5	&	0.84	&	1.04	\\
10$_{3,7}$-9$_{4,6}$	&	5.4904	&	1928.9	&	19	&	21	&	2.6	&	-42	$\pm$	4	&	6	$\pm$	2.4	&	10.3	$\pm$	3.3	&	1.85	$\pm$	1.244	&	-0.87	&	-0.07	&	0.14	\\
\enddata
\end{deluxetable*} 

\startlongtable
\begin{deluxetable*}{l c r c c r cc c c ccc}
\tablecolumns{13}
\tablecaption{Line Parameters of the $-54.5$~\kms\ Component (`H2') in the $\nu_2$=1--0 Transition}
\tablehead{\colhead{Transition}  & \colhead{$\lambda$} 
& \colhead{$E_l$} & \colhead{$g_l$} & \colhead{$g_u$} & \colhead{$A_{ul}$} & \colhead{\vLSR} & \colhead{$\sigma_v$} & \colhead{$N_l$} & \colhead{$W_v$} & \colhead{log$_{10}(\tau_{p, \textrm{thin}})$} & \colhead{log$_{10}(\tau_{p, \textrm{slab}})$} & \colhead{log$_{10}(\beta_0)$}  \\ & \colhead{(\um)} & \colhead{(K)} &  & & \colhead{(s$^{-1}$)} & \colhead{(\kms)} & \colhead{(\kms)}  & \colhead{($\times 10^{15}$~cm$^{-2}$)} & \colhead{(\kms)}  & & & }
\startdata
11$_{2,10}$-10$_{1,9}$	&	5.4248	&	1860	&	21	&	23	&	10.2	&	-55.9	$\pm$	1	&	5.4	$\pm$	1.2	&	2.8	$\pm$	0.4	&	1.83	$\pm$	0.568	&	-0.78	&	0	&	0.19	\\
11$_{6,5}$-11$_{5,6}$	&	5.4521	&	2875.6	&	69	&	69	&	2.4	&	-54.2	$\pm$	1.7	&	5	$\pm$	1.4	&	5.5	$\pm$	0.8	&	0.812	$\pm$	0.24	&	-1.17	&	-0.97	&	-0.68	\\
10$_{1,9}$-9$_{2,8}$	&	5.4862	&	1554.1	&	19	&	21	&	9.5	&	-53.7	$\pm$	1.3	&	5	$\pm$	1.5	&	2.9	$\pm$	0.4	&	1.84	$\pm$	0.61	&	-0.78	&	0.19	&	0.36	\\
9$_{2,7}$-8$_{3,6}$	&	5.5179	&	1447.3	&	51	&	57	&	4.6	&	-54.2	$\pm$	1.1	&	5	$\pm$	1.2	&	6.6	$\pm$	0.8	&	2.08	$\pm$	0.58	&	-0.73	&	0.4	&	0.55	\\
9$_{1,8}$-8$_{2,7}$	&	5.5479	&	1273.9	&	51	&	57	&	8.7	&	-54	$\pm$	1.4	&	5	$\pm$	1.5	&	2.4	$\pm$	0.4	&	1.458	$\pm$	0.544	&	-0.87	&	0.82	&	0.96	\\
10$_{5,6}$-10$_{4,7}$	&	5.5496	&	2274.8	&	63	&	63	&	2.7	&	-54.7	$\pm$	2	&	5	$\pm$	2.4	&	4.3	$\pm$	1.1	&	0.778	$\pm$	0.394	&	-1.19	&	-0.44	&	-0.2	\\
6$_{5,1}$-6$_{4,2}$	&	5.5707	&	1090.1	&	13	&	13	&	1.8	&	-54.9	$\pm$	2.4	&	5	$\pm$	2.8	&	8.8	$\pm$	2.9	&	1.022	$\pm$	0.702	&	-1.04	&	-0.36	&	-0.23	\\
8$_{2,7}$-7$_{1,6}$	&	5.5836	&	1013	&	45	&	51	&	7.9	&	-55.4	$\pm$	3.2	&	7	$\pm$	3.1	&	6.1	$\pm$	1.8	&	3.4	$\pm$	2.162	&	-0.62	&	0.95	&	1.07	\\
\enddata
\end{deluxetable*}

\startlongtable
\begin{deluxetable*}{l c r c c r c cc c c}
\tablecolumns{11}
\tablecaption{Line Parameters of the $-45$~\kms\ Component (`W') in the $\nu_2$=1--0 Transition}
\tablehead{\colhead{Transition}  & \colhead{$\lambda$} 
& \colhead{$E_l$} & \colhead{$g_l$} & \colhead{$g_u$} & \colhead{$A_{ul}$} & \colhead{\vLSR} & \colhead{$\sigma_v$}  & \colhead{$N_l$} & \colhead{$W_v$} & \colhead{log$_{10}(\tau_{p, \textrm{thin}})$}  \\ & \colhead{(\um)} & \colhead{(K)} &  & & \colhead{(s$^{-1}$)} & \colhead{(\kms)} & \colhead{(\kms)} & \colhead{($\times 10^{15}$~cm$^{-2}$)} & \colhead{(\kms)}  & }
\startdata
7$_{1,6}$-6$_{2,5}$	&	5.6921	&	795.4	&	39	&	45	&	6.6	&	-45			&	5	$\pm$	2.5	&	2.9	$\pm$	0.8	&	1.434	$\pm$	0.838	&	-0.44	\\
6$_{0,6}$-5$_{1,5}$	&	5.8304	&	469.9	&	11	&	13	&	10.8	&	-45			&	4.7	$\pm$	1.8	&	2.6	$\pm$	0.7	&	2.488	$\pm$	1.322	&	-0.25	\\
4$_{3,2}$-4$_{2,3}$	&	5.867	&	432.1	&	27	&	27	&	3.2	&	-45			&	5	$\pm$	1.8	&	10.3	$\pm$	1.7	&	2.33	$\pm$	0.868	&	-0.36	\\
5$_{2,4}$-5$_{1,5}$	&	5.8909	&	469.9	&	11	&	11	&	2.5	&	-45			&	5	$\pm$	1.5	&	8.1	$\pm$	1.3	&	1.786	$\pm$	0.526	&	-0.5	\\
4$_{2,2}$-4$_{1,3}$	&	6.0702	&	396.3	&	9	&	9	&	7.7	&	-45			&	5	$\pm$	2.4	&	3.8	$\pm$	0.9	&	2.594	$\pm$	1.236	&	-0.31	\\
2$_{0,2}$-1$_{1,1}$	&	6.1432	&	53.4	&	3	&	5	&	3.8	&	-45			&	5	$\pm$	1.9	&	9.2	$\pm$	1.4	&	4	$\pm$	1.572	&	-0.23	\\
2$_{0,2}$-1$_{1,1}$	&	6.1432	&	53.4	&	3	&	5	&	3.8	&	-45			&	4.6	$\pm$	0.8	&	9.7	$\pm$	0.7	&	4	$\pm$	0.856	&	-0.14	\\
2$_{0,2}$-2$_{1,1}$	&	6.3703	&	136.9	&	5	&	5	&	11	&	-45			&	5	$\pm$	3.3	&	5.2	$\pm$	1.2	&	3.896	$\pm$	2.656	&	0.03	\\
\enddata
\end{deluxetable*}

\startlongtable
\begin{deluxetable*}{l c r c c r c cc c ccc}
\tablecolumns{13}
\tablecaption{Line Parameters of the $-39.5$~\kms\ Component (`H1') in the $\nu_2$=2--1 Transition}
\tablehead{\colhead{Transition}  & \colhead{$\lambda$} 
& \colhead{$E_l$} & \colhead{$g_l$} & \colhead{$g_u$} & \colhead{$A_{ul}$} & \colhead{\vLSR} & \colhead{$\sigma_v$} & \colhead{$N_l$} & \colhead{$W_v$} & \colhead{log$_{10}(\tau_{p, \textrm{thin}})$} & \colhead{log$_{10}(\tau_{p, \textrm{slab}})$} & \colhead{log$_{10}(\beta_0)$}  \\ & \colhead{(\um)} & \colhead{(K)} &  & & \colhead{(s$^{-1}$)} & \colhead{(\kms)} & \colhead{(\kms)}  & \colhead{($\times 10^{15}$~cm$^{-2}$)} & \colhead{(\kms)}  & & & }
\startdata
4$_{4,1}$-3$_{3,0}$	&	5.4373	&	2744.1	&	21	&	27	&	9.8	&	-38.7	$\pm$	1.5	&	4.8	$\pm$	1.2	&	2.1	$\pm$	0.6	&	1.596	$\pm$	0.906	&	-0.85	&	-1.62	&	-1.15	\\
5$_{3,2}$-4$_{2,3}$	&	5.5105	&	2744.7	&	27	&	33	&	6.1	&	-37.6	$\pm$	4.8	&	6	$\pm$	2.8	&	3.5	$\pm$	1.5	&	1.668	$\pm$	1.538	&	-0.92	&	-1.72	&	-1.25	\\
4$_{3,2}$-3$_{2,1}$	&	5.6226	&	2617.1	&	21	&	27	&	8.6	&	-41.3	$\pm$	2.7	&	4.4	$\pm$	2.7	&	1.4	$\pm$	1	&	1.044	$\pm$	1.516	&	-1	&	-1.54	&	-1.08	\\
9$_{2,8}$-8$_{1,7}$	&	5.6247	&	3582.4	&	17	&	19	&	16.1	&	-38.2	$\pm$	3.5	&	4	$\pm$	3.8	&	0.4	$\pm$	0.3	&	0.49	$\pm$	0.838	&	-1.3	&	-2.07	&	-1.57	\\
11$_{0,11}$-10$_{1,10}$	&	5.667	&	3891.4	&	63	&	69	&	27.6	&	-36	$\pm$	2.2	&	5.5	$\pm$	1.6	&	0.9	$\pm$	0.2	&	1.834	$\pm$	0.83	&	-0.82	&	-1.47	&	-0.96	\\
11$_{0,11}$-10$_{1,10}$	&	5.667	&	3891.4	&	63	&	69	&	27.6	&	-40.2	$\pm$	3.1	&	5.2	$\pm$	2	&	1	$\pm$	0.3	&	1.928	$\pm$	1.248	&	-0.77	&	-1.47	&	-0.96	\\
6$_{2,5}$-5$_{1,4}$	&	5.7627	&	2878.3	&	33	&	39	&	10.9	&	-38.5	$\pm$	3.8	&	6	$\pm$	2.2	&	2.4	$\pm$	0.8	&	2.148	$\pm$	1.578	&	-0.78	&	-1.43	&	-0.96	\\
9$_{1,9}$-8$_{0,8}$	&	5.7754	&	3362.5	&	17	&	19	&	25.2	&	-37.4	$\pm$	1.5	&	4.6	$\pm$	1.2	&	0.8	$\pm$	0.2	&	1.546	$\pm$	0.65	&	-0.81	&	-1.69	&	-1.2	\\
\enddata
\end{deluxetable*}

\startlongtable
\begin{deluxetable*}{l c r c c r cc c c c}\label{table:12}
\tablecolumns{11}
\tablecaption{Line Parameters of the $-54.5$~\kms\ Component (`H2') in the $\nu_2$=2--1 Transition}
\tablehead{\colhead{Transition}  & \colhead{$\lambda$} 
& \colhead{$E_l$} & \colhead{$g_l$} & \colhead{$g_u$} & \colhead{$A_{ul}$} & \colhead{\vLSR} & \colhead{$\sigma_v$} & \colhead{$N_l$} & \colhead{$W_v$} & \colhead{log$_{10}(\tau_{p, \textrm{thin}})$}  \\ & \colhead{(\um)} & \colhead{(K)} &  & & \colhead{(s$^{-1}$)} & \colhead{(\kms)} & \colhead{(\kms)}  & \colhead{($\times 10^{15}$~cm$^{-2}$)} & \colhead{(\kms)}  & }
\startdata
11$_{0,11}$-10$_{1,10}$	&	5.667	&	3891.4	&	63	&	69	&	27.6	&	-51.5	$\pm$	3.1	&	7	$\pm$	4	&	0.8	$\pm$	0.3	&	1.582	$\pm$	1.136	&	-1.01	\\
11$_{0,11}$-10$_{1,10}$	&	5.667	&	3891.4	&	63	&	69	&	27.6	&	-55.2	$\pm$	7.2	&	7	$\pm$	11.7	&	0.5	$\pm$	0.5	&	1.058	$\pm$	2.386	&	-1.19	\\
6$_{2,5}$-5$_{1,4}$	&	5.7627	&	2878.3	&	33	&	39	&	10.9	&	-53	$\pm$	6.1	&	7	$\pm$	6.7	&	1.4	$\pm$	1	&	1.352	$\pm$	1.936	&	-1.05	\\
9$_{1,9}$-8$_{0,8}$	&	5.7754	&	3362.5	&	17	&	19	&	25.2	&	-51.9	$\pm$	2.9	&	7	$\pm$	4.1	&	0.7	$\pm$	0.2	&	1.404	$\pm$	0.964	&	-1.07	\\
8$_{1,8}$-7$_{0,7}$	&	5.8322	&	3136.9	&	45	&	51	&	23.9	&	-54	$\pm$	3.5	&	7	$\pm$	5.7	&	0.7	$\pm$	0.4	&	1.47	$\pm$	1.728	&	-1.05	\\
4$_{2,3}$-3$_{1,2}$	&	5.8491	&	2549.6	&	21	&	27	&	9.1	&	-54.5	$\pm$	7.9	&	8	$\pm$	13.3	&	1.1	$\pm$	1.2	&	1.04	$\pm$	2.212	&	-1.27	\\
4$_{2,3}$-3$_{1,2}$	&	5.8491	&	2549.6	&	21	&	27	&	9.1	&	-54.5	$\pm$	7.9	&	8	$\pm$	13.3	&	1.1	$\pm$	1.2	&	1.04	$\pm$	2.212	&	-1.27	\\
6$_{2,5}$-6$_{1,6}$	&	5.9052	&	2938.5	&	39	&	39	&	4.2	&	-53.3	$\pm$	5.5	&	7	$\pm$	8.7	&	1.9	$\pm$	1.4	&	0.648	$\pm$	0.956	&	-1.42	\\
\enddata
\end{deluxetable*}

\end{document}